\documentclass[12pt]{article}

\usepackage{mheck}

\usepackage{amsmath,amssymb,amsfonts}
\usepackage{hyperref}
\usepackage{bbm}
\usepackage{epsfig}
\usepackage[utf8]{inputenc}
\usepackage{color}
\usepackage{cite}
\usepackage{epic}
\usepackage{longtable}
\newcommand{\rem}[1]{} 

\def\Z{\mathbb{Z}}

\def\R{\mathbb{R}}

\def\Hirz[#1]{\mathbbm{F}_{#1}}
\def\o[#1]{\overline{#1}}

\def\S{\mathbb{S}}

\newcommand\ket[1]{\left|#1\right\rangle}
\def\ker{\mbox{ker}}
\def\im{\mbox{im}}

\makeatletter
\newcommand\xleftrightarrow[2][]{%
  \ext@arrow 9999{\longleftrightarrowfill@}{#1}{#2}}
\newcommand\longleftrightarrowfill@{%
  \arrowfill@\leftarrow\relbar\rightarrow}
\makeatother

\institution{OXFORD}{\ Mathematical Institute, University of Oxford Woodstock Road, Oxford, OX2 6GG, UK }
\institution{PERIMETER}{\ Perimeter Institute for Theoretical Physics, 31 Caroline St N, Waterloo, ON N2L 2Y5, Canada }
\institution{WATERLOO}{\ Department of Physics, University of Waterloo, Waterloo, ON N2L 3G1, Canada}

\title{On Mirror Maps for Manifolds of Exceptional Holonomy}

\authors{Andreas P. Braun \worksat{\OXFORD} \footnote{e-mail: {\tt braun@maths.ox.ac.uk}}, Suvajit Majumder \worksat{\OXFORD} \footnote{email: {\tt majumder.suvajit95@gmail.com}} and Alexander Otto \worksat{\PERIMETER,\WATERLOO} \footnote{e-mail: {\tt aotto@perimeterinstitute.ca}}}

\abstract{We study mirror symmetry of type II strings on manifolds with the exceptional holonomy groups $G_2$ and Spin(7). Our central result is a construction of mirrors of Spin(7) manifolds realized as generalized connected sums. In parallel to twisted connected sum $G_2$ manifolds, mirrors of such Spin(7) manifolds can be found by applying mirror symmetry to the pair of non-compact manifolds they are glued from. To provide non-trivial checks for such geometric mirror constructions, we give a CFT analysis of mirror maps for Joyce orbifolds in several new instances for both the Spin(7) and the $G_2$ case. For all of these models we find possible assignments of discrete torsion phases, work out the action of mirror symmetry, and confirm the consistency with the geometrical construction. A novel feature appearing in the examples we analyse is the possibility of frozen singularities. }

\begin{document}

\maketitle
\tableofcontents

\section{Introduction}

The study of string theory on manifolds with the exceptional holonomy groups $G_2$ and $Spin(7)$ from the worldsheet perspective goes back to \cite{Shatashvili:1994zw}. In particular, \cite{Shatashvili:1994zw} (see also \cite{Figueroa-OFarrill:1996tnk}) determined the extension of the worldsheet superconformal algebra for strings propagating on manifolds of exceptional holonomy and pointed out that type II string theories propagating on different $G_2$ (or Spin(7) manifolds) manifolds may result in equivalent physical theories in a phenomenon called $G_2$ (or Spin(7)) mirror symmetry. A necessary condition for any pair of manifolds $M$ and $M^\vee$ to be mirror is that the dimensions of the spaces of exactly marginal operators of the extended $\mathcal{N} = (1,1)$ worldsheet CFTs agree. As shown in \cite{Shatashvili:1994zw}, the number of exactly marginal operators simply equals the number of geometric moduli together with the degrees of freedom associated with the B-field. In the $G_2$ context, this implies the equality of the sum of the second and third Betti numbers for mirrors: 
\begin{equation}\label{eq:g2mirrorcondition}
b^2(M) + b^3(M) =  b^2(M^\vee) + b^3(M^\vee)  \, .
\end{equation}
Likewise, a pair of Spin(7) mirrors $M$, $M^\vee$ must satisfy 
\begin{equation}\label{eq:mirrorcondspin7}
b^2(M) + b^4_-(M) + 1 = b^2(M^\vee) + b^4_-(M^\vee) + 1 \, ,
\end{equation}
where $b^4_-(M)$ denotes the dimension of the space of anti self-dual four-forms. 

This was made explicit for the first time in the context of Joyce orbifolds \cite{joyce1996I,joyce1996II,joyce1996spin7} in \cite{Acharya:1997rh,Acharya:1996fx}. In parallel to the well-studied case of mirror symmetry for type II strings on Calabi-Yau manifolds, where the mirror map can be understood from T-dualities along a calibrated $T^3$ fibration \cite{Strominger:1996it}, mirror maps for type II strings on $G_2$ and Spin(7) manifolds were shown to arise from T-dualities along calibrated $T^3$ or $T^4$ fibrations for some of Joyce's examples in \cite{Acharya:1997rh,Acharya:1996fx,Gaberdiel:2004vx,Chuang:2004th}, see also \cite{Papadopoulos:1995da}.

More recently, \cite{MR2024648,MR3109862,Corti:2012kd} have given a construction of $G_2$ manifolds as twisted connected sums (TCS) by gluing appropriate pairs of asymptotically cylindrical (acyl) Calabi-Yau threefolds $X_\pm$ (times a circle $\mathbb{S}^1_\pm$ on each side). Mirror maps for TCS $G_2$ manifolds were found in \cite{Braun:2017ryx,Braun:2017csz}, where it was shown that applying a mirror map to both of the acyl Calabi-Yau threefolds, together with T-dualities on the product circles, leads to another TCS $G_2$ manifold which satisfies \eqref{eq:g2mirrorcondition}. This map can be described as being the result of performing four T-dualities along a calibrated $T^4$ fibration, so that it maps type IIA (IIB) strings to IIA (IIB) strings. The $T^4$ fibre in question is understood as the product of the Strominger-Yau-Zaslow (SYZ) fibres of the acyl Calabi-Yau threefolds times the circles $\mathbb{S}^1_\pm$. 

Interestingly, TCS $G_2$ manifolds allow a second class of mirror maps satisfying \eqref{eq:g2mirrorcondition}, in which only one of the two acyl Calabi-Yau threefolds $X_\pm$ is exchanged for its mirror \cite{Braun:2017csz}. In case one of the acyl Calabi-Yau threefolds carries an elliptic fibration, this mirror map can likewise be understood from three T-dualities along a $T^3$ fibre\footnote{As discussed in \cite{joyce2000compact}, it is not expected on general grounds for $G_2$ manifolds to have such fibrations, so it might be better to speak of such $G_2$ manifolds as admitting a limit in their moduli space in which an associative $T^3$ collapses.}, i.e. this duality must map type IIA strings to type IIB or vice versa. For compact Calabi-Yau manifolds which are hypersurfaces or complete intersections in toric varieties, mirror families have an elegant combinatorial construction using pairs of reflexive polytopes \cite{Batyrev:1994hm}. As shown in \cite{Braun:2016igl,Braun:2017ryx}, a completely analogous construction exists for acyl Calabi-Yau threefolds in terms of pairs of dual tops, which makes it possible to give large numbers of concrete examples of $G_2$ mirrors. An intriguing feature of both classes of mirror maps is that they sometimes map smooth geometries to singular ones \cite{Braun:2017csz}. This is analogous to mirror symmetry in the context of K3 surfaces \cite{Aspinwall:1994rg}, where the presence of the B-field prevents the occurrence of extra massless states.

Calibrated torus fibrations are not just interesting in the study of mirror symmetry of Calabi-Yau, $G_2$ or Spin(7) manifolds, but also feature in the duality between M-Theory and heterotic string theory. The duality in seven dimensions between M-Theory on K3 surfaces and heterotic string theory on a three-torus $T^3$ with a flat connection can be used fibrewise to find examples of lower-dimensional dualities. This was exploited for the SYZ fibration of Calabi-Yau threefolds to find examples of heterotic duals of M-Theory on TCS $G_2$ manifolds in \cite{Braun:2017uku} (see also \cite{Halverson:2014tya,Halverson:2015vta,Guio:2017zfn} for explorations of M-Theory on TCS $G_2$ manifolds). Furthermore, $T^4$ fibrations also play a crucial role for the determination of the superpotential of M-Theory on TCS $G_2$ manifolds. As argued in \cite{Braun:2018fdp,Acharya:2018nbo}, there is a large class of associative submanifolds of TCS $G_2$ manifolds which appear as sections of precisely the coassociative $T^4$ fibration relevant for mirror maps. 

Applying the M-Theory duality map to heterotic strings on TCS $G_2$ manifolds yields M-Theory duals on Spin(7) manifolds with a decomposition similar to TCS, called generalized connected sum (GCS) in \cite{Braun:2018joh}. GCS Spin(7) manifolds are glued from two non-compact manifolds with the holonomy groups SU(4) and $G_2$, so that it is tempting to exploit this structure to construct candidates of mirror manifolds by using mirrors for the building blocks. This is an analogue of the strategy used in \cite{Braun:2017ryx,Braun:2017csz} for TCS $G_2$ manifolds and we will follow a similar path to define Spin(7) mirrors in the present work, and show that they indeed satisfy \eqref{eq:mirrorcondspin7}. 

Given geometric constructions for mirrors of $G_2$ and Spin(7) manifolds, it becomes an interesting question if these can be recovered using worldsheet methods. This has been accomplished for a few of the examples of Joyce. These are resolutions of orbifolds, so that they can be treated from first principles in string theory \cite{Acharya:1997rh,Gaberdiel:2004vx}. Furthermore, these geometries can also be decomposed as twisted connected sums. As shown in \cite{Braun:2017csz}, the two complimentary approaches result in the same mirror maps and identify the same $T^3$ and $T^4$ fibrations. Furthermore, it has subsequently been shown \cite{Fiset:2018huv} that the $G_2$ mirror maps of \cite{Braun:2017ryx,Braun:2017csz} are associated with non-trivial automorphisms of the extended superconformal algebra of the worldsheet theory for TCS $G_2$ manifolds.

It is a central motivation of the present work to enlarge the class of models where a geometrical construction of mirrors for $G_2$ and Spin(7) manifolds can be compared with results obtained from the worldsheet, and we complete this task for several new examples.

Treating new examples of orbifolds, and linking the associated mirror maps with a geometry requires several steps to be completed. Crucially, the definition of string theory on the orbifolds we are considering involves an assignment of discrete torsion phases \cite{vafa1986modular}. Possible assignments of discrete torsion phases can be constrained by the requirement of modular invariance for the partition function. A short introduction to how such constraints arise and can be analysed is presented in Appendix \ref{sect:introdtandmodinv}. Different assignments of discrete torsion will in turn lead to a different spectrum of RR ground states, which via the map to cohomology \cite{Witten:1982im} signals the correspondence to topologically different smoothings of the orbifold geometry. Mirror symmetry can change the discrete torsion phases, so that it also associates CFTs on different smooth geometries. The action of mirror maps can be found by providing a free-field realization of the extended superconformal algebra, and finding automorphisms of this algebra which are induced by a sequence of T-dualities. Finally, to establish a link to the TCS mirror maps the orbifolds in question must be described as twisted connected sums.

In the $G_2$ examples studied in \cite{Gaberdiel:2004vx}, the possible assignments of discrete torsion precisely match the different resolutions found in \cite{joyce1996II} (`example 4'). Furthermore, the description of such models as TCS $G_2$ manifolds in \cite{Braun:2017csz} allowed for a straightforward determination of torsion in the homology group $H^3(M_k,\mathbb{Z})$, which precisely matches with the discrete torsion phases in the orbifold string theory, as expected from \cite{Aspinwall:1995rb,1998math......9072G}. 

In Section \ref{G2}, we will complete these tasks for a set of models which are free quotients of the orbifolds considered in \cite{Gaberdiel:2004vx, Braun:2017csz}, they have been first presented as `example 5' in \cite{joyce1996II}.  As we will see, these orbifolds are an example the 'extra-twisted' connected sums described in \cite{2015arXiv150502734C}. Curiously, not all resolutions which have been constructed in \cite{joyce1996II} are realized by the set of consistent assignments of discrete torsion phases. 

In Section \ref{sect:newspin7example}, we consider Joyce orbifolds which can be smoothed to manifolds with holonomy Spin(7). Some Joyce orbifolds and their mirror maps have been previously from the worldsheet perspective in \cite{Acharya:1997rh,Chuang:2004th}. As only submanifolds of real dimension four can be calibrated for Spin(7) manifolds, such mirror maps must be associated with four T-dualities along a calibrated $T^4$ fibration. In Section \ref{sect:newspin7example} of the present work we study two such examples, which first appeared as `example 1' and `example 2' in \cite{joyce1996spin7}, as well the action of mirror symmetry on such geometries. 

Finally, Section \ref{sect:oldspin7example} presents a general exposition of how the GCS construction can be used to define mirror maps for Spin(7) manifolds. While our construction is motivated by the identification of a fibration by $T^4$, the resulting check of \eqref{eq:mirrorcondspin7} for the mirror geometries holds independently. We then proceed to describe how the examples studied in Section \ref{sect:newspin7example} are decomposed as a GCS and verify that our geometric construction of GCS Spin(7) mirrors precisely agrees with the worldsheet results. 

The appendices contain a brief introduction to discrete torsion phases and modular invariance for strings on orbifolds, as well technical details of the examples we are treating.


\section{A $G_2$ Example}\label{G2}

In this section we consider the following example from \cite{joyce1996II} (`example 5'). It is based on a quotient of $T^7$ under the group $\Gamma = \Z_2^4$, with an action on the coordinates
\begin{equation}\label{eq:orbig2examplesigma2}
\begin{array}{c|ccccccc}
& X^1 & X^2 & X^3 & X^4 & X^5 & X^6 & X^7 \\
\hline 
\alpha & + & + & + & - & - & - & - \\
\beta & + & - & - & + & + & -\tfrac12 & - \\
\gamma & - & + & - & + & - & + &- \tfrac12 \\
\sigma_2 & + & \tfrac12 & + & \tfrac12 & + & + & +
\end{array}\, .
\end{equation}
Here  $X^i \sim X^i + 1$ giving $T^7$, and $-$ indicates that the corresponding coordinate is sent to minus itself, while $\tfrac12$ indicates a shift $X \rightarrow X + \tfrac12$, and $-\tfrac12$ is a shorthand for $X \rightarrow -X + \tfrac12$.

\subsection{Smoothing of the Orbifold}\label{sect:g2smoothing}

Let us first analyse the topological properties of the smooth limit(s) obtained from this orbifold by resolving the singularities as in \cite{joyce1996II} (see also \cite{joyce2000compact}). In order to do that, we need to first evaluate the fixed point set under the action of the orbifold group. 

The only group elements of $\Gamma$ which act non-freely and hence give rise to singularities are $\alpha$, $\beta$, and $\gamma$. Each of these three elements fixes 16 T$^3$s, which are then further identified under the action of the rest of the orbifold group. In case of the $\alpha$ sector, the action of $\langle \beta,\gamma,\sigma_2 \rangle$ is free, resulting in two orbits of 8 T$^3$ each. For the $\beta$ sector, we have the same result as the $\alpha$ sector as these two sectors are isomorphic, up to the permutation of indices and fixed point labels. The analysis is a bit different for the $\gamma$ case. Here we have a further $\mathbb{Z}_2$ identification under the action of $\alpha\beta$. This leads to 8 orbits consisting of 2 T$^3/\mathbb{Z}_2$ each. The action of $\sigma_2$ just reduces the fundamental domain for $X^2$ and $X^4$ (two of the extended directions in the $\gamma$ fixed T$^3$s). Hence, the $\gamma$ action has a singular set consisting of 8 T$^3/\mathbb{Z}_2$. 

\begin{center}
\begin{tabular}{|c|c|c|c|}
    \hline
           & singular set & elements in orbit under $\Gamma$ & singular set in quotient\\
    \hline
   $\alpha$  &  16 $T^3$ & 8$T^3$ & 2 $T^3$\\
   $\beta$   & 16 $T^3$ & 8$T^3$ & 2 $T^3$\\
   $\gamma$  & 16 $T^3$ & 2$T^3/\mathbb{Z}_2$ & 8 $T^3/\mathbb{Z}_2$\\
   \hline
\end{tabular}    
\end{center}

The net contribution to the singular set is obtained by adding up the separate contributions from the different orbifold actions, giving us a total of 4 T$^3$, and 8 T$^3/\mathbb{Z}_2$. With the singular set at hand, we are now ready to compute the Betti numbers for the different smooth limits obtained by resolving the singularities present in our orbifold. First of all, the Betti numbers for the orbifold are given by $(b^2,b^3)(T^7/\Gamma)=(0,7)$. Each of the singularities is locally modelled on T$^3\times \mathbb{C}^2/\{\pm 1\}$ or (T$^3\times \mathbb{C}^2/\{\pm 1\})/\langle\alpha\beta\rangle$. The contribution to the Betti numbers from resolving these singularities is: 
\begin{equation}
\begin{aligned}
\delta (b^2,b^3)(T^3)&=(1,3)\\
\delta (b^2,b^3)(T^3/\mathbb{Z}_2)&=(1,1)\,\, \text{or}\,\, (0,2)
\end{aligned}
\end{equation}
The two choices for the latter case come from different possible smoothings. Denoting the compact smooth $G_2$ manifold obtained by making the first choice $k$ times by $\hat{M}_k$ we find
\begin{equation}
\begin{aligned}
 b^2(\hat{M}_k) &= 4+k \\
 b^3(\hat{M}_k) &= 35 - k
\end{aligned}
\end{equation}
for $k=0 \cdots 8$. 

\subsection{Constraints on Discrete Torsion}

Let us redo the analysis by studying the string theory living on the orbifold $T^7/\mathbb{Z}^4_2$. For such string theories, we have the additional degree of freedom to switch on discrete torsion phases consistent with modular invariance constraints \cite{vafa1986modular}. We then exploit the isomorphism between the Ramond-Ramond sector ground states and the target space cohomology to find the Betti numbers for the resulting resolution(s).

We shall begin by studying the twisted sectors corresponding to various elements of the orbifold group. If the orbifold element has fixed points in its action on the parent manifold, the corresponding sector can then be further decomposed into sub-sectors localised at those fixed points. In general, one would start by figuring out which of the sectors can actually have discrete torsion signs. For this, we need to write down the representation matrices for the orbifold elements in a basis of highest weight states in a particular sector. However, as our main aim is to obtain the cohomology of the resulting orbifold smoothings, we can get away by working with a few of the twisted sectors only: ones that contribute to the ground state spectrum (and hence the cohomology), and ones that are needed to constrain the discrete torsion signs for the ground state contributing sectors. 

In our case, the ground state contributions come from the $\alpha$, $\beta$, and $\gamma$ sectors, as all the remaining group elements act on at least one direction as $+\frac{1}{2}$, implying that the oscillator modes are half-integer in that direction. Aside from these sectors, we need to consider $\alpha\beta$, $\sigma_2$, and $\sigma_2\alpha\beta$ sectors in order to figure out the modular trace constraints in the $\alpha$, $\beta$, and $\gamma$ sectors. Let us then find which of the above mentioned sectors have discrete torsion signs. The details of this analysis can be found in Appendix \ref{G2aplha}.

{\boldmath $\alpha$}\textbf{-} and {\boldmath$\beta$} \textbf{-sector:} There are no discrete torsion choices in these sectors.

{\boldmath$\gamma$}\textbf{-sector:} In the $\gamma$ sector, we have the first instance of discrete torsion. We find that there are two orbifold generators with possible discrete torsion signs in their irreducible representations: $\alpha\beta$ and $\sigma_2$. The corresponding phases are denoted by $\epsilon_{f_{\gamma}}(\alpha\beta)$, $\epsilon_{f_{\gamma}}(\sigma_2)$ respectively, where $f_{\gamma}=1,2,..,8$ labels the different irreducible representations/orbits. 

As reviewed in Appendix A, one can constrain the discrete torsion signs by implementing modular invariance of the partition functions. For that, we utilize the orbifold elements with non-zero traces in the above representations and probe the twisted sector corresponding to those elements. These are given by the group elements $\alpha\beta$, $\sigma_2$, and $\sigma_2\alpha\beta$.

{\boldmath$\alpha\beta$}\textbf{-sector:} In this sector, we have discrete torsion signs appearing in the representation matrix corresponding to the $\gamma$ generator:

\begin{equation}
    \gamma|_{\mathcal{H}_{\alpha\beta}^{f_{\alpha\beta}}}=\begin{pmatrix}\epsilon^1_{f_{\alpha\beta}}(\gamma)\mathbb{I}_{4\times 4}&0\\
    0&\epsilon^2_{f_{\alpha\beta}}(\gamma)\mathbb{I}_{4\times 4}\end{pmatrix}
\end{equation}
Invariance of the partition function under the S transformation implies
\begin{equation}
\begin{split}
        \sum_{f_{\alpha\beta},i}\epsilon^i_{f_{\alpha\beta}}(\gamma)&=\sum_{f_{\gamma}=1}^8\epsilon_{f_{\gamma}}(\alpha\beta)\label{traceconstraintg-ab}
\end{split}
\end{equation}
where $f_{\alpha\beta}=1,2,3,4$ labels the irreducible representations, and i=1,2 labels the two signs in a given irreducible representation.

{\boldmath$\sigma_2$}\textbf{-sector:} As shown in Appendix \ref{G2gamma}, discrete torsion arises in the representation of the element $\gamma$, $\gamma|_{\mathcal{H}_{\sigma_2}}=\epsilon_{\sigma_2}(\gamma)\mathbb{I}_{4\times4}$. We can now use the S transform relations to connect this discrete torsion sign with $\epsilon_{f_{\gamma}}(\sigma_2)$ from the $\gamma$ twisted sector as follows:
\begin{equation}\label{sigma2-constraint1}
    8\epsilon_{\sigma_2}(\gamma)=\sum_{f_{\gamma}=1}^8\epsilon_{f_{\gamma}}(\sigma_2)
\end{equation}

{\boldmath$\sigma_2\alpha\beta$}\textbf{-sector:} Analogous to the $\sigma_2$ sector, discrete torsion arises in the representation of the element $\gamma|_{\mathcal{H}_{\sigma_2\alpha\beta}^{f_{\sigma_2\alpha\beta}}}=\epsilon_{f_{\sigma_2\alpha\beta}}(\gamma)\mathbb{I}_{8\times 8}$, as is expected from modular constraint requirements. Using modular invariance of partition function under the S-transformation gives
\begin{equation}\label{sigma2-constraint2}
\begin{split}
2\sum_{f=1}^4\epsilon_{f_{\sigma_2\alpha\beta}}(\gamma)&=\sum_{f=1}^8\epsilon_{f_{\gamma}}(\sigma_2)\epsilon_{f_{\gamma}}(\alpha\beta)
\end{split}
\end{equation}

\subsection{Cohomology of smoothings}\label{sect:constandhomologyg2}

We are ready to derive the cohomology of the possible resolutions of the orbifold by exploiting its isomorphism with the RR ground states. Let us first consider the untwisted sector, $\mathcal{H}_e$. We have Majorana-Weyl spinors $\psi^i$ corresponding to the 7 bosonic coordinates $X^i$, i=1,...,7. We can generate the RR ground states by acting on the vacuum $\ket{0}$ with the creation operators built out of the zero modes, $\psi^i_{+}=(\psi_0^i+i\tilde{\psi_0^i})/2$. The orbifold invariant set of RR ground states is given by:
\begin{equation}\label{untwisted2}
\ket{0}; \hspace{10 pt} \psi^i_+\psi^j_+\psi^k_+\ket{0}; \hspace{10pt} \psi^a_+\psi^b_+\psi^c_+\psi^d_+\ket{0}; \hspace{10pt} \psi^1_+...\psi^7_+\ket{0}
\end{equation}
with the following the triples and the 4-tuples of indices:
\begin{equation}
(i,j,k)\in \{(1,2,3),(1,4,5),(1,6,7),(2,4,6),(2,5,7),(3,5,7),(3,4,6)\}
\end{equation}
\begin{equation}
(a,b,c,d)\in \{(4,5,6,7),(2,3,6,7),(2,3,4,5),(1,3,5,7),(1,3,4,6),(1,2,4,6),(1,2,5,7)\}
\end{equation}
Now the isomorphism between RR ground states and the target space cohomology allows us to use the following identification:
\begin{equation}\label{isomorphism-statement}
\psi_+^{i_1}...\psi_+^{i_n}\ket{0}\simeq dX^{i_1}\wedge...\wedge dX^{i_n}
\end{equation}
As such, we can use the list of invariant states from \eqref{untwisted2} to get the Betti number contribution from the untwisted sector:
\begin{equation}\label{untwisted7D}
\begin{split}
b^0_u=b^7_u=1\\
b^3_u=b^4_u=7 \, 
\end{split}
\end{equation}
For $G_2$ manifolds, the only unfixed, non-trivial Betti numbers are the second and third ones. Now on, we will only mention contributions to those.

Aside from the untwisted sector, the smoothings for the orbifold will also get contribution to the Betti numbers from the twisted sectors with RR ground states, i.e. $\alpha$, $\beta$, and $\gamma$. Now, we shall evaluate such twisted sector contributions.

Let us look at the contribution from $\alpha$ sector. Firstly, following in the same line of argument as in \cite{Gaberdiel:2004vx}, we should identify the vacuum state in this sector with the exceptional divisor $\Sigma$ resolving the particular singularity (a 2-form in this case). Then, we list the invariant states in the $\alpha$ sector and use a similar isomorphism statement as in the untwisted sector case (in \eqref{isomorphism-statement}) to get the Betti numbers contribution. 

The list of $\Gamma$ invariant RR ground states in the $\alpha$ twisted sector is built on the highest weight states $\ket{0}^{f_{\alpha}}_{\alpha}$ with vanishing momentum and winding modes, by acting with the creation operators built from the Majorana Weyl fermions as in the untwisted sector. The only difference is that we only have zero modes along $X^1$, $X^2$, and $X^3$. Here the label $f_{\alpha}=1,2$ enumerates the two irreducible representations corresponding to $X^5={0,1/2}$ as in Appendix B. They are given by:
\begin{equation}\label{twisted-2}
\ket{0}^{f_{\alpha}}_{\alpha}; \hspace{10pt} \psi^i_+\ket{0}^{f_{\alpha}}_{\alpha}; \hspace{10pt} \psi^{i_1}_+\psi^{i_2}_+\ket{0}^{f_{\alpha}}_{\alpha}: \hspace{10pt} \psi^{k_1}_+\psi^{k_2}_+\psi^{k_3}_+\ket{0}^{f_{\alpha}}_{\alpha}
\end{equation}
where i$\in$\{1,2,3\}; (i$_1$,i$_2$)$\in$ \{(1,2),(1,3),(2,3)\}; (k$_1$,k$_2$,k$_3$)$\in$\{(1,2,3)\}. The isomorphism statement can be again used to get the following identification:
\begin{equation}
    \psi_+^{i_1}...\psi_+^{i_n}\ket{0}^{f_{\alpha}}_{\alpha}\simeq dX^{i_1}\wedge...\wedge dX^{i_n}\wedge\Sigma_{f_{\alpha}}
\end{equation}
where $\Sigma_{f_{\alpha}}$ represents the exceptional divisor arising due to the resolution of the singularity corresponding to the orbit ($\sim$ irreducible representation) labelled by $f_{\alpha}$. Now, we can read off the Betti numbers $\delta b^i_{\alpha}$ from the above list as:
\begin{equation}\label{Bettialpha}
\begin{split}
    \delta b^2_{\alpha}=2 \cdot 1=2\\
    \delta b^3_{\alpha}=2 \cdot 3=6
\end{split}
\end{equation}
where the factor of 2 comes from the index $f_{\alpha}$. As the $\beta$ sector is isomorphic to the $\alpha$ sector, we find the same contributions there.

For the $\gamma$-sector, things are different as there are discrete torsion signs present. Firstly, we need to solve for the trace relations in \eqref{traceconstraintg-ab}, \eqref{sigma2-constraint1}, \eqref{sigma2-constraint2},  that constrain the discrete torsion phases arising in this sector. 
There are two distinct cases corresponding to the sign of $\epsilon_{\sigma_2}$. For $\epsilon_{\sigma_2}=1$, \eqref{sigma2-constraint1} implies that the only possible solution is $\epsilon_{f_{\gamma}}(\sigma_2)=1$ for all $f_{\gamma}$. Also from \eqref{sigma2-constraint2} we get: 
\begin{equation}
2\sum_{f_{\sigma_2\alpha\beta}=1}^4\epsilon_{f_{\sigma_2\alpha\beta}}(\gamma)=\sum_{f_{\gamma}=1}^8\epsilon_{f_{\gamma}}(\alpha\beta) \, .
\end{equation}
Note that we are forced by this relation to have an even number of positive discrete torsion signs $\epsilon_{f_{\gamma}}(\alpha\beta)$. Thus, we can re-label the fixed point indices such that:
\begin{equation}\label{eq:numbermustbeeven}
\epsilon_{f_{\gamma}}(\alpha\beta)=\epsilon_{8-f_{\gamma}}(\alpha\beta),\hspace{10pt} f_{\gamma}=1,2,3,4
\end{equation}
Now let us list the orbifold invariant states as before. For $\epsilon_{f_{\gamma}}(\alpha\beta)=\epsilon_{f_{\gamma}}=1$, they are:
\begin{equation}\label{gamma+}
\ket{0}_{\gamma}^{f_{\gamma}}; \hspace{10pt} \psi_+^6\ket{0}_{\gamma}^{f_{\gamma}}; \hspace{10pt}\psi_+^{2}\psi_+^4\ket{0}_{\gamma}^{f_{\gamma}}; \hspace{10pt} \psi_+^{2}\psi_+^4\psi_+^6\ket{0}_{\gamma}^{f_{\gamma}}
\end{equation}
where $\ket{0}_{\gamma}^{f_{\gamma}}$ is the highest weight state representing the irreducible basis for the different orbits indexed by $f_{\gamma}$. On the other hand, for $\epsilon_{f_{\gamma}}(\alpha\beta)=-\epsilon_{f_{\gamma}}=-1$, the orbifold invariant states are:
\begin{equation}\label{gamma-}
\psi_+^{2}\ket{0}_{\gamma}^{f_{\gamma}};\hspace{10pt} \psi_+^4\ket{0}_{\gamma}^{f_{\gamma}}; \hspace{10pt} \psi_+^{2}\psi_+^6\ket{0}_{\gamma}^{f_{\gamma}}; \hspace{10pt} \psi_+^{4}\psi_+^6\ket{0}_{\gamma}^{f_{\gamma}};
\end{equation}
For the Betti number contributions, we need the state-cohomology identification statement as before. In this case, the highest weight state is again mapped to a 2-form for either case of the discrete torsion sign of $\epsilon_{f_{\gamma}}(\alpha\beta)$. This is because the exceptional divisor in both cases corresponds to a 2-form. Using the same identification as in the $\alpha$ and $\beta$ sectors, we can read off the Betti numbers from the list of invariant states in \eqref{gamma+}, and \eqref{gamma-} as:
\begin{equation}
\begin{split}
        &\delta b^2_{\gamma}=1; \hspace{10pt}\delta b^3_{\gamma}=1,\hspace{5pt} \text{iff}\hspace{5pt}\epsilon_{f_{\gamma}}(\alpha\beta)=1\\
        &\delta b^2_{\gamma}=0; \hspace{10pt}\delta b^3_{\gamma}=2,\hspace{5pt} \text{iff}\hspace{5pt}\epsilon_{f_{\gamma}}(\alpha\beta)=-1 
\end{split}
\end{equation}
Thus, the resulting Betti number contribution for 2$l$ positive signs of $\epsilon_{f_{\gamma}}(\alpha\beta)$ is:
\begin{equation}\label{Bettigamma}
\begin{split}
    &\delta b^2_{\gamma}=2l \cdot 1+(8-2l) \cdot 0=2l\\
    &\delta b^3_{\gamma}=2l \cdot 1+(8-2l) \cdot 2=16-2l
\end{split}
\end{equation}
for $l = 0 \cdots 4$. These are all the contributions from the twisted sectors as the rest of the sectors all have at least one extended direction with half-integer modes. Let the resultant smoothing for $2l$ positive signs of $\epsilon_{f_{\gamma}}(\alpha\beta)$ be called $\hat{M}_{2l}$. The net Betti number for $\hat{M}_{2l}$ can be obtained by adding up the contributions from the three twisted sectors $\alpha,\beta,\gamma$ along with that coming from the untwisted sector. Using \eqref{untwisted7D}, \eqref{Bettialpha} and \eqref{Bettigamma}, we get:
\begin{equation}\label{case1}
\begin{aligned}
& b^2(\hat{M}_{2l})=b^2_u+\delta b^2_{\alpha}+\delta b^2_{\beta}+\delta b^2_{\gamma}=0+2+2+2l=&4+2l\\
& b^3(\hat{M}_{2l})=b^3_u+\delta b^3_{\alpha}+\delta b^3_{\beta}+\delta b^3_{\gamma}=7+6+6+(16-2l)=&35-2l
\end{aligned}
\end{equation}
for $l = 0 \cdots 4$. Although every single one of these models corresponds to one of the resolutions discussed in Section \ref{sect:g2smoothing}, not all of the possible geometries are realized. This ultimately stems from the constraint that the number of discrete torsion sign in \eqref{eq:numbermustbeeven} must be even. It would be interesting to understand this mismatch better, but as our main interest is in the action of mirror maps on these models we leave such an investigation to future work.  

Now let us move onto the other case where we have $\epsilon_{\sigma_2}=-1$. Here, we don't get any states from the $\gamma$-twisted sector as no invariant states can be constructed with $\epsilon_{f_{\gamma}}(\sigma_2)=-1$. The Betti numbers for the partially smoothed solution, say $\hat{N}$, is then obtained by omitting the contribution from the $\gamma$ sector in our previous computation in \eqref{case1}:
\begin{equation}\label{case2}
\begin{split}
    & b^2(\hat{N})=b^2_u+\delta b^2_{\alpha}+\delta b^2_{\beta}=0+2+2=4\\
    & b^3(\hat{N})=b^3_u+\delta b^3_{\alpha}+\delta b^3_{\beta}=7+6+6=19
\end{split}
\end{equation}
What we have here is a scenario where the orbifold singularities could only partially be smoothed, but some are frozen. In particular, all of the singularities located at the $\gamma$ fixed points must be left intact. The freezing of singularities is a well-known phenomenon for strings on orbifolds in the presence of discrete torsion \cite{Vafa:1994rv}, but it is, to the authors knowledge, the first time it has been observed for $G_2$ orbifolds.

\subsection{Mirror Symmetry}\label{sect:g2mirrorcft}

Let us now have a look at the realisation of mirror symmetry in our G2 example through T-duality transformations on multiple suitably chosen coordinates. 

The extended chiral algebra for a string moving on a compact G2 manifold consists of a $\mathcal{N}=1$ superconformal algebra generated by the bosonic stress tensor T along with its fermionic counterpart G, extended by currents ($\phi$,X) of conformal dimensions ($h_{\phi},h_X$)=(3/2,2) and their superpartners K,M. Their free field representation is
\begin{equation}\label{eq:g2algebragens}
\begin{aligned}
    T_{G_2}&=1/2\sum_{j=1}^7:\partial X^j\partial X^j:-1/2\sum_{j=1}^7:\psi^j\partial\psi^j:\\
    G_{G_2}&=\sum_{j=1}^7\psi^j\partial X^j\\
    X_{G_2}&=-\psi^4\psi^5\psi^6\psi^7-\psi^2\psi^3\psi^6\psi^7-\psi^2\psi^3\psi^4\psi^5-\psi^1\psi^3\psi^5\psi^7+\psi^1\psi^3\psi^4\psi^6+\psi^1\psi^2\psi^5\psi^6+\psi^1\psi^2\psi^4\psi^7\\
    \Phi_{G_2}&=\psi^1\psi^2\psi^3+\psi^1\psi^4\psi^5+\psi^1\psi^6\psi^7+\psi^2\psi^4\psi^6-\psi^2\psi^5\psi^7-\psi^3\psi^4\psi^7-\psi^3\psi^5\psi^6\\
    M_{G_2}&=[G,X]\\
    K_{G_2}&=\{G,\Phi\}
\end{aligned}
\end{equation}
A mirror automorphism of this algebra is given by:
\begin{equation}\label{mirrorG2}
    \begin{array}{c|cccccc}
& T_{G2} & G_{G2} & \Phi_{G2} & X_{G2} & K_{G2} & M_{G2} \\
\hline 
\text{mirror}_{G_2} & + & + & - & + & - & + \\
\end{array}
\end{equation}
Defining two sets of triples of coordinate indices as follows:
\begin{equation}
\begin{aligned}
    \mathcal{I}_3^+&=\{(1,6,7),(2,4,6),(3,5,6)\}\\
    \mathcal{I}_3^-&=\{(1,2,3),(1,4,5),(2,5,7),(3,4,7)\}
\end{aligned}\, ,
\end{equation}
one can check that the elements $\mathcal{I}_3^\pm$ are precisely those T-duality triples that generate the mirror automorphism of eq. \eqref{mirrorG2} on the right-movers (left chiral algebra stays invariant). As the present model is a quotient of the one considered in \cite{Gaberdiel:2004vx}, it should come as no surprise that these are the same triples which were found there (after an appropriate relabelling of coordinates). 

The resultant action of these composite T-dualities on the discrete torsion phases can be understood through the following line of argument. If we focus on the representation of the $\alpha\beta$ generator in the $\gamma$-twisted sector (knowing that it has discrete torsion phases $\epsilon_{f_{\gamma}}(\alpha\beta)$ arising in its representation), we observe a need for flipping all or none of the discrete torsion signs upon action of these transformations. On the other hand, the signs corresponding to the $\sigma_2$ generator, i.e. $\epsilon_{f_{\gamma}}(\sigma_2)$ do not change. In the $\gamma$-twisted sector, the RR zero modes surviving are labelled the indices (2,4,6). Now $\alpha\beta$ flips the signs of coordinate labels (2,4) while $\sigma_2$ does not flip any of the three labels. As a result, we can write the elements in terms of the RR zero modes:
\begin{equation}
\begin{aligned}
    \alpha\beta&=\psi_0^2\psi_0^4\tilde{\psi}_0^2\tilde{\psi}_0^4\epsilon_{f_{\gamma}}(\alpha\beta)\\
    \sigma_2&=\epsilon_{f_{\gamma}}(\sigma_2)
\end{aligned}\, .
\end{equation}
The above expressions imply that under the set of T-transformations in $\mathcal{I}_3^+$, $\epsilon_{f_{\gamma}}(\alpha\beta)$ should remain the same; while for $\mathcal{I}_3^-$, $\epsilon_{f_{\gamma}}(\alpha\beta)$ should flip signs irrespective of the fixed point label f. On the other hand, the $\sigma_2$ representation matrix remaining invariant under the triples of T-dualities implies that $\epsilon_{f_{\gamma}}(\sigma_2)$ has to remain invariant as well. This implies that
\begin{equation}
\begin{aligned}
\mathcal{I}_3^+ :& \hat{M}_{2l} \rightarrow \hat{M}_{2l}  \\
\mathcal{I}_3^- :& \hat{M}_{2l} \rightarrow \hat{M}_{8-2l}  
\end{aligned}\, .
\end{equation}
Furthermore, the case with frozen singularities, represented by \eqref{case2}, should be considered self-mirror. Note that all of these maps take type IIA strings to type IIB strings and vice versa because of the odd number of fermionic modes being T-dualized. 

We can combine any two of the above transformations to get a new transformation which T-dualizes along 4 of the 7 coordinates. These transformations split into $\mathcal{I}_4^+$, and $\mathcal{I}_4^-$, with $\mathcal{I}_4^+$ keeping the discrete torsion signs $\epsilon_{f_{\gamma}}(\alpha\beta)$ intact and $\mathcal{I}_4^-$ inverting all of them
\begin{equation}
\begin{aligned}
    \mathcal{I}_4^+&=\{(1,2,4,7),(1,3,5,7),(2,3,4,5)\}\\
    \mathcal{I}_4^-&=\{(2,3,6,7),(4,5,6,7),(1,2,5,6),(1,3,4,6)\}
\end{aligned}\, .
\end{equation}
In particular, these act as
\begin{equation}
\begin{aligned}
\mathcal{I}_4^+ :& \hat{M}_{2l} \rightarrow \hat{M}_{2l}  \\
\mathcal{I}_4^- :& \hat{M}_{2l} \rightarrow \hat{M}_{8-2l}  
\end{aligned}\, .
\end{equation}
Now as there are even number of fermionic modes, the mirror maps corresponding to these transformations do not alter the sign of GSO projection, mapping type IIA and IIB strings to IIA and IIB respectively.

\subsection{Realization as a TCS}

Let us now discuss how the orbifold treated above and its smoothings $\hat{M}_k$ can be described as a extra twisted connected sum. Twisted connected sum $G_2$ manifolds are constructed as \cite{MR2024648,Corti:2012kd}
\begin{equation}\label{eq:TCS}
M =\left( X_+ \times \S^1_+\right) \hspace{.2cm} \#  \hspace{.2cm} \left( X_- \times \S^1_-\right) \, ,
\end{equation}
for a pair of asymptotically cylindrical Calabi-Yau threefolds $X_\pm$ which enjoy fibration by K3 surfaces $S_\pm$. In their asymptotic regions $X_\pm$ are approximated (metrically) by the product of $S_\pm$ and a cylinder, i.e. $S_\pm \times \S^1_\mp \times I$ for an interval $I$. The gluing to $M$ is then done by identifying those asymptotic regions by a diffeomorphism which induces am appropriate hyper-K\"ahler rotation on the K3 surfaces $S_\pm$. 

The example $\hat{M}_k$ discussed above is not a TCS $G_2$ manifold. However, it can be constructed as a quotient of another $G_2$ manifold $M_k$ by the free involution generated by $\sigma_2$. The realization of $M_k$ as a TCS has been discussed in detail in \cite{Braun:2017csz}. The upshot of this analysis is that the K3 fibres $S_\pm$ of both acyl Calabi-Yau threefolds $X_\pm$ are given as a (smoothing) of $T^4/\gamma$ with the coordinates $(X^1,X^3,X^5,X^7)$. The acyl Calabi-Yau threefold $X_+$ can be described as $T^5 \times \R / \langle \gamma, \alpha \rangle$, with $X^6$ parametrizing the non-compact direction and $(X^1,X^3,X^5,X^7,X^4)$ parametrizing the $T^5$. Hence the base of the K3 fibration on $X_+$ has coordinates $(X^4,X^6)$ and the coordinate along $\S^1_+$ is $X^2$. Likewise, $X_-$ can be described as $T^5 \times \R / \langle \gamma, \beta \rangle$, also with $X^6$ parametrizing the non-compact direction. Now, however the base of the K3 fibration on $X_-$ has coordinates $(X^2,X^6)$ and the coordinate along $\S^1_-$ is $X^4$, as appropriate for gluing these two acyl CYs to a TCS $G_2$ manifold. 

The freely acting involution $\sigma_2$ hence only acts on the coordinates $X^2,X^4,X^6$ along the base $S^3$ of the K3 fibration apparent in the TCS decomposition of $M_k$. In particular, it acts by shifting the coordinates on both of the $S^1$ factors in the decomposition of $S^3$ appearing in the TCS construction, where it is glued from two solid tori.\footnote{This is in fact the same freely-acting involution of $S^3$ which is naturally found by representing $S^3$ as
\begin{equation}\label{eq:S3inC2}
|z_1|^2 + |z_2|^2 = 1 
\end{equation}
and acting with 
\begin{equation}\label{eq:sigma2actsonzi}
 \sigma_2 : \hspace{.3cm}
 \begin{array}{cc}
  z_1 & \rightarrow - z_1 \\
  z_2 & \rightarrow - z_2
 \end{array}
\end{equation}
In this presentation, the decomposition into two solid tori can be seen by solving \eqref{eq:S3inC2} using the parametrization
\begin{equation}
\begin{aligned}
z_1 & = e^{i\phi} \cos(\eta) \\
z_2 & = e^{i\psi} \sin(\eta)
\end{aligned}
\end{equation}
Here, $\phi$ and $\psi$ have the range $0\cdots 2\pi$ and $\eta$ parametrizes the interval $0\cdots \pi$, so that we can see $S^3$ decomposed as a being glued from the two solid tori described by the above for $0\leq \eta \leq \pi/2$ and $\pi/2 \leq \eta \leq \pi$. In this parametrization, \eqref{eq:sigma2actsonzi} is simply the map 
\begin{equation}\label{eq:sigma2actsonphipsi}
 \sigma_2 : \hspace{.3cm}
 \begin{array}{cc}
  \phi & \rightarrow \phi + \pi \\
  \psi & \rightarrow \psi+ \pi
 \end{array}
\end{equation}
which is nothing but the half-shift induced by $\sigma_2$ in \eqref{eq:orbig2examplesigma2} again. } The further quotient by $\sigma_2$ turns this into an example of an \emph{extra-twisted connected sum} as defined in \cite{2015arXiv150502734C}. This construction differs from the usual TCS construction \cite{MR2024648} in two regards: first of all, the $G_2$ manifold $M$ is glued from two pieces
\begin{equation}\label{eq:etcs}
M = V_+ \hspace{.2cm} \#_\vartheta \hspace{.2cm} V_- \, , 
\end{equation}
where $V_\pm$ are free quotients of $X_\pm \times \S^1_\pm$ by $\Z_2^s$ for acyl Calabi-Yau threefolds $X_\pm$. In the neck region in which $X_\pm$ asymptote to $S_\pm \times \S^1_\mp \times I$ for K3 surfaces $S_\pm$, the quotient must purely act by shifts on the $\S^1$ factors, so that
\begin{equation}
V_\pm \setminus \kappa_\pm =  S_\pm \times I \times \left( \S^1_+ \times \S^1_- \right)/\Z_2^s
\end{equation}
for compact subsets $\kappa_\pm$. Note that $\left( \S^1_+ \times \S^1_- \right)/\Z_2^s$ is still a two-torus. The second difference is that the gluing now involves an angle $\vartheta$ with which these tori are identified, together with an appropriate altering of the hyper-K\"ahler rotation acting on the K3 surfaces $S_\pm$ to keep the canonically defined $G_2$ forms $\Phi_\pm$ invariant. In the example discussed here, the `trivial' choice $\vartheta = \pi/2$ appears, which means that the two $\S^1$s are simply swapped, as in the standard TCS construction. 

\subsection{TCS Mirror Map}

We can now describe a mirror map in this context, which is found by a slight generalization of the approach of \cite{Braun:2017ryx,Braun:2017csz}. There, the central idea was to construct a mirror of a TCS $G_2$ manifold by applying mirror symmetry to either both, or to one of the two acyl Calabi-Yau threefolds in the TCS decomposition \eqref{eq:TCS}. The present example of a extra-twisted connected sum is constructed as a free quotient $\hat{M}_k$ of the TCS $G_2$ manifolds $M_k$, which can also be described as acting separately on the acyl Calabi-Yau threefold $X_\pm$. 

Omitting the action of $\sigma_2$ in \eqref{eq:orbig2examplesigma2}, the TCS decomposition and the action of the TCS mirror map on the resulting $G_2$ manifolds $M_k$ was described in \cite{Braun:2017csz}. The result is that both $X_\pm$ are such that 
\begin{equation}
\begin{aligned}
|K_\pm| & = 4 \\
h^{2,1}(Z_\pm) &= 4 \\
|N_\pm| & = 10 \\
|N_+ \cap N_-| & = k 
\end{aligned}\, ,
\end{equation}
where $K = \ker \left( H^{1,1}(X_\pm) \rightarrow H^{1,1}(S_\pm) \right)$, $N_\pm = \im \left( H^{1,1}(X_\pm) \rightarrow H^{1,1}(S_\pm) \right)$ and $Z_\pm$ is the compactification of $X_\pm$ found by gluing in a single K3 fibre (see \cite{MR3109862} for more details). The topology of the resulting $G_2$ manifold is then determined from 
\begin{equation}
\begin{aligned}
 b^2 &= |K_+| + |K_-| + |N_+ \cap N_-| \\
b^2+b^3 &= 23 + 2(|K_+| + |K_-|) + 2(h^{2,1}(Z_+) +h^{2,1}(Z_-) )
\end{aligned}\, . 
\end{equation}

Orbifolding $X_\pm \times \S^1_{\pm}$ by the freely acting $\Z_2^s$ produces a 7-manifold $V_\pm$ with the holonomy group $SU(3) \rtimes \Z_2^s$ (`barely $G_2$'). In the present case, both the elliptic fibrations on $X_\pm$ and the SYZ fibrations on $X_\pm$ become fibrations by $T^2$ and $T^3$ on $V_\pm$. Repeating the analysis of \cite{Braun:2017csz} then motivates to consider two types of mirrors of $\hat{M}_k$:
\begin{equation}
\hat{M}^\vee =   V_+^\vee \hspace{.2cm} \# \hspace{.2cm} V_-^\vee
\end{equation}
associated with applying four T-dualities, as well as
\begin{equation}
\hat{M}^\wedge =   V_+ \hspace{.2cm} \# \hspace{.2cm} V_-^\vee
\end{equation}
associated with applying three T-dualities. 

Given the data of $X_\pm$, the Betti numbers of the free quotients $M_k/ \Z_2^s$ are given by
\begin{equation}
 \begin{aligned}
b^2 &= |K_+^e| + |K_-^e| + |N_+ \cap N_-| \\
b^2+b^3 &= 23 + 2(|K_+^e| + |K_-^e|) + 2(h^{2,1}_e(Z_+) +h^{2,1}_e(Z_-) )
\end{aligned}\, . 
\end{equation}
where $^e$ and $_e$ indicates taking the even subspace under the involution $\Z_2^s$. Note that the group $\Z_2^s$ does not act on the K3 fibres of $X_\pm$, so that $N_\pm$ are unchanged. In the present case, we have that
\begin{equation}
\begin{aligned}
|K_\pm^e| & = 2 \\
h^{2,1}_e(Z_\pm) &= 2 \\
|N_\pm| & = 10 \\
|N_+ \cap N_-| & = k 
\end{aligned}\, ,
\end{equation}
so that we recover
\begin{equation}
\begin{aligned}
b^2(\hat{M}) &= 4+k \\ 
b^3(\hat{M}) &= 35-k \\ 
\end{aligned}\, .
\end{equation}

We are now ready to discuss the action of the TCS mirror maps. The fact that $X_\pm$ are self-mirror indicates that the same is true for $V_\pm$. As in the analysis of the Joyce orbifold in \cite{Braun:2017csz}, the only non-trivial ingredient in the mirror construction is given by $N_+ \cap N_-$. As $\sigma_2$ does not act on the K3 fibres at all, we can just quote the result of the analysis of \cite{Braun:2017csz}: whereas $|N_+ \cap N_-| = k$ for $\hat{M}^\vee$, $|N_+ \cap N_-| = 8 - k$ for $\hat{M}^\wedge$. We hence find that 
\begin{equation}
 \begin{aligned}
 b^2(\hat{M}_k^\vee) &= 4 + k \\
b^3(\hat{M}_k^\vee) &= 35 - k 
 \end{aligned}
\end{equation}
whereas
\begin{equation}
 \begin{aligned}
b^2(\hat{M}_k^\wedge) &= 12 - k\\
b^3(\hat{M}_k^\wedge) &= 27 + k
 \end{aligned}\, .
\end{equation}
As in \cite{Braun:2017csz}, we can associate $\hat{M}^\vee$ with the image under $\mathcal{I}_4^+$ and $\hat{M}^\wedge$ with the image under $\mathcal{I}_3^-$.


\section{Spin(7) Examples}\label{sect:newspin7example}

In this section we will analyse two $T^8/\mathbb{Z}_2^4$ orbifolds in which the generators $\a$, $\beta$, $\gamma$ and $\delta$ of $\Gamma = \mathbb{Z}_2^4$ act on the $T^8$ coordinates as \cite{joyce1996spin7}:
\begin{equation}\label{eq:orbispin7example}
\begin{array}{c|cccccccc}
& X^1 & X^2 & X^3 & X^4 & X^5 & X^6 & X^7 & X^8 \\
\hline 
\alpha & - & - & - & - & + & + & + & + \\
\beta & + & + & + & + & - & - & - & - \\
\gamma & -\frac{c_1}{2} & -\frac{c_2}{2} & + & + & -\frac{c_5}{2} & -\frac{c_6}{2} & + & +\\
\delta & -\frac{d_1}{2} & + & -\frac{d_3}{2} & + & -\frac{d_5}{2} & + & -\frac{d_7}{2} & +
\end{array}\, .
\end{equation}
where $c_i,\,d_i\in\{0,1\}$. The two cases we will study have $c_i=(1,1,1,1)$ and $d_i=(0,1,1,1)$ (`Example 1'), and $c_i=(1,0,1,0)$ and $d_i=(0,1,1,1)$ (`Example 2'). Once again, $-1/2$ is shorthand for $X^i \rightarrow 1/2-X^i$


\subsection{Smoothing of the Orbifold}\label{smoothingofspin7}

First, let us review the resolution of the $T^8/\mathbb{Z}_2^4$ orbifolds above into a family of compact $Spin(7)$ manifolds as described in \cite{joyce1996spin7}. In total, the Betti numbers will receive a contribution from the orbifold itself, as well as contributions form the resolutions of the singularities at the fixed points. Before resolution, the cohomology consists of the classes of $T^8$ which are invariant under the group $\mathbb{Z}_2^4$. In either example, a simple calculation shows $b^0=b^8=1$ and $b^4=14$. For example, $dX^1\wedge dX^2 \wedge dX^3$ is not invariant, while $dX^1\wedge dX^2\wedge dX^3\wedge dX^4$ is. In particular, the 14 4-forms are precisely the 14 elementary ones appearing in the 4-form on $\mathbb{R}^8$ that $Spin(7)$ is defined to leave invariant (see e.g. equation (1) of \cite{joyce1996spin7}).

Now consider example 1, we would like to understand its resolutions and so must first understand its singular set. In this case, it is clear that only $\alpha$, $\beta$, $\gamma$, $\delta$ and $\alpha\beta$ have fixed points, as all other combinations involve $X^i\rightarrow X^i+1/2$ for some $i$, which has no fixed points. So, the singular set consists of 5 sectors, $S_\alpha$, $S_\beta$, $S_\gamma$, $S_\delta$ and $S_{\alpha\beta}$.

For each single generator, the fixed points correspond to 16 $T^4$'s for the directions unchanged by the group action (For example, $\a$ has fixed points $X^i\in\{0,1/2\}$ for $i=1..4$ and $X^i$ free for $i=5..8$). For $\alpha\beta$, the only composite element with fixed points, they are the 256 points $X^i\in\{0,1/2\}$ for $i=1..8$. However, we must also consider the action of the other group generators on a given set of fixed points. For the case of $\alpha$, $\beta$ acts on the fixed $T^4$'s by $-1$, giving 16 $T^4/\{\pm1\}$'s. The subgroup $\langle \gamma, \delta \rangle$ acts freely, and so groups $S_\alpha$ into $4$ orbits of $T^4/\{\pm1\}$'s. In a similar way, we see that the set $S_\beta$ also consists of 4 copies of $T^4/\{\pm1\}$. For the $\gamma$-fixed points, $\langle \alpha, \beta, \delta \rangle$ acts freely and so we find that the singular set gets reduced to 2 $T^4$'s, and we apply the same reasoning to find that $S_\delta$ is also 2 sets of $T^4$'s. Lastly, $\langle \gamma, \delta \rangle$ acts freely on the $\alpha\beta$ fixed points, grouping the 256 points into 64 (here, $\langle \alpha \beta \rangle$ acts freely). In summary, the fixed set for example 1 looks like:

\begin{center}
\begin{tabular}{ |c|c|c|c| } 
 \hline
& singular set & elements in orbit under $\Gamma$ & singular set in quotient \\  \hline
  $\a$      & 16 $T^4$'s & 4 $T^4/\{\pm1\}$'s & 4 $T^4/\{\pm1\}$'s \\
  $\beta$   & 16 $T^4$'s & 4 $T^4/\{\pm1\}$'s & 4 $T^4/\{\pm1\}$'s \\
  $\gamma$  & 16 $T^4$'s & 8 $T^4$'s & 2 $T^4$'s \\
  $\delta$  & 16 $T^4$'s & 8 $T^4$'s & 2 $T^4$'s \\
  $\a\beta$ & 256 points & 4 points & 64 points\\
 \hline
 
 \hline
\end{tabular}
\end{center}

Lastly, we consider the neighbourhood of each singular point. For $S_\a$ and $S_\beta$, the neighbourhoods of the fixed loci are a total of 8 copies of $(T^4/\{\pm1\})\times (B^4_\zeta/\{\pm1\})$ for a $4$-ball $B^4_\zeta$. In the language of Proposition 3.1.1 in \cite{joyce1996spin7}, these come from type (ii) singularities, which upon resolution increases $b^1$ by 1 and $b^4$ by 6. For $S_\gamma$ and $S_\delta$, the neighbourhoods consist of 4 $T^4\times (B^4_\zeta/\{\pm1\})$'s, which arise from type (i) singularities, and increase $b^2$ by 1, $b^3$ by 4 and $b^4$ by 6. Lastly, the neighbourhoods of $S_{\a\beta}$ consist of 64 $(B^4_\zeta/\{\pm1\})\times (B^4_\zeta/\{\pm1\})$'s, which are of type (iii) and increase $b^4$ by 1. In total, we find that upon resolution the non-trivial Betti numbers of the manifold are:
\begin{equation}
\begin{aligned}
(b^2,\,b^3,\,b^4)=(12,\,16,\,150)\,\,.
\end{aligned}
\end{equation}

Let us now consider example 2. In this case, all sectors except the $\gamma$-fixed points contribute the same singularities, and thus Betti numbers, as example 1. However, for $\gamma$, $\a\delta$ now acts trivially. Starting from a neighbourhood $T^4\times B_\zeta^4$, the $\gamma$ action converts this to $T^4\times (B_\zeta^4/\{\pm1\})$, but in fact the action of $\a\delta$ turns this into the neighbourhood of a singularity of type (iv) \cite{joyce1996spin7}. The action of the rest of the group then orders these 16 type (iv) singularities into 4 orbits. In summary, the singular set of example 2 is given by
\begin{center}
\begin{tabular}{ |c|c|c|c| } 
 \hline
& singular set & elements in orbit under $\Gamma$ & singular set in quotient\\  \hline
  $\a$      & 16 $T^4$'s & 4 $T^4/\{\pm1\}$'s & 4 $T^4/\{\pm1\}$'s \\
  $\beta$   & 16 $T^4$'s & 4 $T^4/\{\pm1\}$'s & 4 $T^4/\{\pm1\}$'s \\
  $\gamma$  & 16 $T^4$'s & 4 $T^4$'s & 4 $T^4$'s \\
  $\delta$  & 16 $T^4$'s & 8 $T^4$'s & 2 $T^4$'s \\
  $\a\beta$ & 256 points & 4 points & 64 points\\
 \hline
 
 \hline
\end{tabular}
\end{center}
In this case, the singularities induced by the action of $\gamma$ may be resolved in two inequivalent ways, one increasing $b^2$ by 1, $b^3$ by 2 and $b^4$ by 2, and the other increasing $b^3$ by 2 and $b^4$ by 4. As a result, including the contribution from the unresolved orbifold and letting $j\in\{0..4\}$ be the number of type (iv)'s we resolve in the first way, we find a family of 5 $Spin(7)$ manifolds with Betti numbers given by:
\begin{equation*}
    (b^2,b^3,b^4)=(10+j,16,154-2j)\,,\quad j\in\{0..4\}\,\,.
\end{equation*}


\subsection{Discrete Torsion and Cohomology}

In this section we will recover the resolution of the orbifold by studying the associated CFT. Once again, following a similar analysis used in \cite{Gaberdiel:2004vx} we will relate the distribution of the allowed discrete torsion signs (which will mostly be derived in the appendix) to the cohomology of the orbifold resolutions.


\subsubsection{Example 1}\label{sect:spin7example1}

We begin with Example 1, where $c_i=(1,1,1,1)$ and $d_i=(0,1,1,1)$. The cohomology of the smooth limit of the orbifold must be related to the ground states of RR ground states in each twisted sector. In particular, this means we are studying zero momentum states, which only arise in twisted sectors associated with group elements acting non-freely. Thus we only need to consider contributions from the generators $\alpha$, $\beta$, $\gamma$, $\delta$, as well as $\a\beta$. Using the analysis in Appendix \ref{app:spin7}, we find the following results for the discrete torsion phases:

\textbf{$\a$ and $\beta$ sectors}: In each case the 16 twisted sectors get ordered into 4 sets of 4 labelled by numbers $f_\a$ and $f_\beta$, both running from 1 to 4 and thus contributing 4 ground states each. In each sector, we have a discrete torsion sign $\e_{f_\a}(\beta)$ for the action of $\b$ on the $\a$ states and a sign $\e_{f_\beta}(\alpha)$ in the other direction. In this case modular invariance requires the distribution of these signs within the $\a$ and $\beta$ sectors to be the same:
\begin{equation}
\sum_{f_\a=1}^4\e_{f_\a}(\beta)= \sum_{f_\beta=1}^4\e_{f_\beta}(\alpha)\,\,.
\end{equation}

\textbf{$\gamma$ and $\delta$ sectors}: In these sectors, the 16 fixed points get organized into 2 sets of 8, however there is no discrete torsion sign available and so we only have 2 ground states to work with.

\textbf{$\a\beta$ sector}: In this sector, we have 256 fixed points organized into 64 sets of 4 labelled by a number $f_{\a\beta}$ running from 1 to 64. We have a discrete torsion degree of freedom for both the action of $\a$ and $\b$ on these states, which in fact must be equal and we call it $\e_{f_{\a\beta}}(\a,\beta)$. We can then label the fixed points such that $f_{\a\beta}=f_\a+4k$ for $k=0,..,15$ and get $\e_{f_\a+4k}(\a,\beta)=\e_{f_\a}$. In other words, the 64 possible signs get grouped into 4 sets of 16 signs which must all be equal, and how we distribute such signs across these sets of 16 must correspond to how we distribute the signs in the $\a$ and $\beta$ sectors:
\begin{equation}
 \sum_{f_{\a\beta}=1}^{64}\e_{f_{\a\beta}}(\beta)=16\sum_{f_\a=1}^4\e_{f_\a}
\end{equation}

With this in mind, we can now determine the contribution to the cohomology from the different twisted sectors. We begin first with the $\gamma$ and $\delta$ sectors, for which there are no discrete torsion phases. In either case, the 16 fixed point/twist fields get grouped into 2 sets of 8 states. Following the methods in \cite{Gaberdiel:2004vx}, we need to find linear combinations of these 8 states in each of the two sectors (we call such a combination $\ket{0,0;f_g}_g$, for $f_g=1,2$ and $g=\gamma$ or $\delta$) such that when acted on by the raising operators $\psi_+^i$ (where $i\in I_\gamma=\{3,4,7,8\}$ for $\gamma$ and $i\in I_\delta =\{2,4,6,8\}$ for $\delta$) we obtain $\Z_2^4$ invariant states. In doing so we find states of the following form:
\begin{equation*}
    \ket{0,0;f_g}_{g}\,,\quad \psi_+^i\ket{0,0;f_g}_{g}\,,\quad \psi_+^i\psi_+^j\ket{0,0;f_g}_{g}\,,\quad \psi_+^i\psi_+^j\psi_+^k\ket{0,0;f_g}_{g}\,,\quad \psi_+^i\psi_+^j\psi_+^k\psi_+^l\ket{0,0;f_g}_{g}\,\,.
\end{equation*}
Where $i,j,k,l\in I_g$ and $g=\gamma$ or $\delta$. In total, there is one state of the first form, 4 of the second, 6 of the 3rd, 4 of the 4th and one of the 5th for each of $g=\gamma,\delta$. Thus, identifying $\ket{0,0;f}_g\simeq$ a 2 form as in \cite{Gaberdiel:2004vx} and $\psi_+^i\simeq dX^i$ we find the contribution (recalling that the distribution of signs within each sector must be the same):
\begin{equation}
\begin{aligned}
\delta b^2_g & = 2  \\
\delta b^3_g & = 8  \\
\delta b^4_g & = 12  \\
\end{aligned}
\label{eq:yd}
\end{equation}
for $g=\gamma$ or $\delta$.

The $\alpha$, $\beta$ and $\alpha\b$ sectors are more interesting, as they involve the choice of a discrete torsion sign. Beginning with $\a$ (the $\b$ sector is identical), we recall that our 16 twist fields were grouped into 4 sets of 4 and follow a similar procedure to above. Provided we choose the appropriate signs between states in one orbit, we find that when we choose $\e_{f_\alpha}(\beta)$ or $\e_{f_\beta}(\a)$ to be +1 we get states:
\begin{equation*}
    \ket{0,0;f_g}_{g}\,,\quad \psi_+^i\psi_+^j\ket{0,0;f_g}_{g}\,,\quad \psi_+^i\psi_+^j\psi_+^k\psi_+^l\ket{0,0;f_g}_{g}\,\,,
\end{equation*}
Where $i,j,k,l\in I_\a=\{5,6,7,8\}$ or $I_\beta=\{1,2,3,4\}$ and $g=\a$ or $\b$. So, we have 1 state with no oscillators, 6 with 2, and 1 with 4. When we take it to be $-1$, we get states with 1 and 3 $\psi_+^i$'s with $i$ taking values in the same possible index set. So, we would find 1 state with no oscillators, 4 with 1, and 4 with 3. If $k$ is the number of signs we take to be +1, and we make the same identifications of states with forms as before, we find the contribution:
\begin{equation}
\begin{aligned}
\delta b^2_g & = k  \\
\delta b^3_g & = 16-4k  \\
\delta b^4_g & = 6k  \\
\end{aligned}
\end{equation}
for $g=\a$ or $\beta$ and $k=0,...,4$. 

For $\a\b$, the 256 fixed points were organized into 64 sets of 4. There are no oscillators to use as raising operators, so we only have states $\ket{0,0;f_{\a\beta}}_{\a\b}$. However, there are 2 interesting things to note here. Firstly, this state should not be interpreted as a 2 form, but rather a 4 form (loosely, we can think of these as products of $\a$ and $\b$ singularities, and so the associated form as a product of 2 forms, giving a 4 form). Secondly, recall that choosing one of the $\a$ or $\beta$ discrete torsion phases to be -1 meant choosing 16 of the $\e_{f_{\a\beta}}(\a,\beta)$'s to be -1 for consistency. When we do so, because $\a$ and $\b$ act diagonally on the states within the 64 orbits, we find we cannot create any invariant linear combination. Thus, if $k$ is again the number of positive signs we find that this sector contributes:
\begin{equation}
\begin{aligned}
\delta b^4_{\a\beta} & = 16k  \\
\end{aligned}
\end{equation}
We must also include the contribution from the untwisted sectors, which arises from the states $\prod_i(\psi_+^i)^{l_i}\ket{0}$ where $\ket{0}$ is the untwisted ground state and each $l_i\in\{0,1\}$. In particular, we identify $\ket{0}$ with the constant 0-form, each $\psi_+^i$ with $dX^i$, and take $\mathbb{Z}_2^4$ invariant states. As expected, we find that they are in one to one correspondence with the invariant classes in $H^i(T^8/\mathbb{Z}_2^4)$, and so the total Betti numbers are, after adding up each $\delta^p_g$ for $g=\a$, $\beta$, $\gamma$, $\delta$ and $\a\beta$:
\begin{equation}\label{eq:bettispin7ex1}
    (b^2,b^3,b^4)(M_k)=(10,2k+4,48-8k)\,,\quad k\in\{0..4\}\,\,.
\end{equation}

Rather interestingly, this does not actually completely agree with the result in \cite{joyce1996spin7}! Rather, it only reduces to the expected result when $k=4$. At first, we may naively believe we have found a set of other new resolutions, but on returning to the previous analysis we find that this is not quite the case. Recall that when $k<4$, some of the 64 $\a\b$-twisted sectors could not provide us with states, as they were not invariant. We then interpret this as saying we are \textit{not} resolving 16 of the associated $\a\beta$ singularities. Thus, when $k<4$, we do not actually have a complete resolution - rather, we have only a partial resolution of the orbifold and the remaining singularities are frozen due to the presence of discrete torsion. 


\subsubsection{Example 2}\label{sect:spin7example2}

Now we analyze example 2, where $c_i=(1,0,1,0)$ and $d_i=(0,1,1,1)$. This model is very similar to example 1, the full analysis for this sector is done in Appendix \ref{app:spin7}. The only crucial difference being that the $\gamma$, $\a\delta$ and $\a\delta\gamma$ sectors all receive a discrete torsion sign - however, modular invariance constrains them so that we may set the distribution of these signs must be the same within each sector:
\begin{equation}
4\sum_{f_{\gamma}=1}^4\e_{f_\gamma}(\delta)
=4\sum_{f_{\a\delta\gamma}=1}^4\e_{f_{\a\delta\gamma}}(\delta)
=2\sum_{f_{\a\delta}=1}^8\e_{f_{\a\delta}}(\gamma)\,\,. 
\end{equation}
Once again, of these 3 only the $\gamma$ sector will have zero momentum states to contribute to the cohomology, and  the contributions $\delta b^p_g$ for $g=\a$, $\beta$, $\delta$ and $\a\beta$ are the same as in example 1.

The 16 fixed points of the $\gamma$ sector are organized into 4 sets of 4 labelled by a number $f_\gamma\in\{1,..,4\}$, and so we build states from linear combinations of states within the orbits, which we call $\ket{0,0;f_\gamma}_\gamma$. We then fill out the ground states with $\psi_+^i$ for $i=3,4,7,8$. The discrete torsion phases come from the actions of $\delta$, $\a\delta$ and $\a\delta\gamma$ , but consistency of the representation requires $\e_{f_{\gamma}}(\delta)=\e_{f_{\gamma}}(\a\delta)=\e_{f_{\gamma}}(\a\delta\gamma)\equiv \e_{f_{\gamma}}$. When we choose $\e_{f_{\gamma}}=+1$, we find that the only invariant states are:
\begin{equation*}
    \ket{0,0;f_\gamma}_\gamma\,,\quad\psi_+^3\ket{0,0;f_\gamma}_\gamma\,,\quad\psi_+^8\ket{0,0;f_\gamma}_\gamma\,,\quad\psi_+^3\psi_+^8\ket{0,0;f_\gamma}_\gamma\,,\quad\psi_+^4\psi_+^7\ket{0,0;f_\gamma}_\gamma\,\,,
\end{equation*}
whereas when $\e_{f_{\gamma}}=-1$ we find the states:
\begin{gather*}
    \psi_+^4\ket{0,0;f_\gamma}_\gamma\,,\quad\psi_+^7\ket{0,0;f_\gamma}_\gamma\,,\quad\psi_+^3\psi_+^4\ket{0,0;f_\gamma}_\gamma\,,\quad\psi_+^3\psi_+^7\ket{0,0;f_\gamma}_\gamma\,,\quad\\
    \quad\psi_+^4\psi_+^8\ket{0,0;f_\gamma}_\gamma\,,\quad\quad\psi_+^7\psi_+^8\ket{0,0;f_\gamma}_\gamma\,\,.
\end{gather*}
Letting $j$ be the number of $\e_{f_{\gamma}}$ which are equal to $1$, we find that the contribution from this sector to the cohomology of the associated resolution is:
\begin{equation}
\begin{aligned}
\delta b^2_\gamma & = j  \\
\delta b^3_\gamma & = 8  \\
\delta b^4_\gamma & = 16-2j  \\
\end{aligned}
\end{equation}
and thus the total cohomology is given by:
\begin{equation}
    (b^2,b^3,b^4)(M_{j,k})=(2+2k+j,48-8k,42+28k-2j)\,\,,
\end{equation}
where $M_{j,k}$ represents the space we obtain after the different resolutions of the orbifold, parametrized by $j$ and $k$. When $k=4$ we get:
\begin{equation}
    (b^2,b^3,b^4)(M_{j,4})=(10+j,16,154-2j)\,\,.
\end{equation}
which are the Betti numbers found in \cite{joyce1996spin7}. Once again, a choice of $k< 4$ means we cannot create certain states in the $\a\b$ sector and some of the singularities are frozen by discrete torsion. 


\subsection{Mirror Symmetry}

In this section, we will briefly discuss mirror symmetry for the smooth $Spin(7)$ manifolds obtained in example 2. Example 1 can be treated analogously, with the result that all of the (partially or fully resolved) models obtained there are self-mirror.

The starting point is the $Spin(7)$ superconformal algebra, the generators of which can be obtained from the generators $(T_{G_2},G_{G_2},\Phi_{G_2},X_{G_2},K_{G_2},M_{G_2})$, \eqref{eq:g2algebragens}, of the $G_2$ algebra as follows:
\begin{align*}
    T&=T_{G_2}+\frac{1}{2}:\partial X^8\partial X^8:-\frac{1}{2}:\psi^8\partial\psi^8:\,\,,\\
    G&=G_{G_2}+:\psi^8\partial X^8:\,\,,\\
    X&=X_{G_2}+\Phi_{G_2}\psi^8+\frac{1}{2}\psi^8\partial\psi^8\,\,,\\
    M&=[G,X]=\partial X^8 \Phi_{G_2}-K_{G_2}-M_{G_2}+\frac{1}{2}\partial^2 X^8\psi^8-\frac{1}{2}\partial X^8\partial \psi^8\,\,.
\end{align*}
The algebra these operators satisfy is worked out in \cite{Shatashvili:1994zw}. Written this way, it is easy to work out the analogue of the $G_2$ mirror automorphism by combining the $G_2$ map:
\begin{equation}
    (T_{G_2},G_{G_2},\Phi_{G_2},X_{G_2},K_{G_2},M_{G_2})\rightarrow (T_{G_2},G_{G_2},-\Phi_{G_2},X_{G_2},-K_{G_2},M_{G_2})
\end{equation}
together with a T-duality along $X^8$. Doing so, we see that this combination maps the algebra directly back on to itself. Using  combinations of the 3-direction T-dualities found in \cite{Gaberdiel:2004vx} combined with T-duality along $X^8$, we can then explicitly realize this automorphism as such a duality. We find that the following 7 combinations generate our $Spin(7)$ automorphism:
\begin{equation*}
    \{(2,4,6,8),\,(2,3,5,8),\,(1,2,7,8),\,(1,3,6,8),\,(1,4,5,8),\,(3,4,7,8),\,(5,6,7,8)\}\,\,.
\end{equation*}
Another option is to use an automorphism that leaves the $G_2$ algebra invariant, without affecting the terms involving $X^8$ or $\psi^8$. This can be done using the combinations of 4 $T$-dualities found in \cite{Gaberdiel:2004vx}, which leave the $G_2$ algebra invariant and so map the $Spin(7)$ algebra on to itself once again. The following 7 index sets generate an automorphism of the $Spin(7)$ algebra in this way:
\begin{equation*}
    \{(1,2,5,7),\,(1,4,6,7),\,(3,4,5,6),\,(2,4,5,7),\,(2,3,6,7),\,(1,2,5,6),\,(1,2,3,4)\}\,\,.
\end{equation*}

We would now like to see how these maps act on any discrete torsion signs. To do so, we construct representations of the operators in terms of the $\psi_0^i$ and $\tilde{\psi}_0^i$'s. Our first focus is on the $\gamma$ sector, for which it is the $\a\delta$ parity that we are interested in. A quick calculation gives:
\begin{equation*}
    \a\delta|_{\mathcal{H}_\gamma^{f_{\gamma}}}=\frac{1}{4}\psi_0^4\psi_0^7\tilde{\psi}_0^4\tilde{\psi}_0^7\epsilon_{f_\gamma}(\delta)\,\,.
\end{equation*}
Next, we want to consider the representation of $\a$ in the $\beta$ sector, $\beta$ in the $\a$ sector, and $\a$ and $\b$ in the $\a\b$ sector. We find, for $\beta$ in the $\a$ sector:
\begin{equation}
    \beta|_{\mathcal{H}_\a^{f_{\a}}}=\frac{1}{16}\psi_0^5\psi_0^6\psi_0^7\psi_0^8\tilde{\psi}_0^5\tilde{\psi}_0^6\tilde{\psi}_0^7\tilde{\psi}_0^7\cdot\epsilon_{f_\a}(\beta)\, .
    \label{eq:bona}
\end{equation}
and for $\a$ acting in the $\beta$ sector:
\begin{equation}
    \a|_{\mathcal{H}_\beta^{f_{\beta}}}=\frac{1}{16}\psi_0^1\psi_0^2\psi_0^3\psi_0^4\tilde{\psi}_0^1\tilde{\psi}_0^2\tilde{\psi}_0^3\tilde{\psi}_0^4\cdot\e_{f_\beta}(\a)\, .
    \label{eq:aonb}
\end{equation}
In the $\a\beta$ sector we have no zero modes, and so there is no representation of $\a$ or $\beta$ in terms of the $\psi$'s (i.e. it can be represented by $\a|_{\mathcal{H}_{\a\beta}^{f_{\a\beta}}}=\beta|_{\mathcal{H}_\a^{f_{\a}}}=\e_{f_{\a\beta}}(\a,\beta)$). Applying the T-dualities, we find that the 14 possible combinations group into two sets:
\begin{align*}
    I_+=&\{(2,3,5,8),\,(1,3,6,8),\,(3,4,7,8),\,(1,4,6,7),\,(2,4,5,7),\,(1,2,5,6)\}\,\,.\\
    I_-=&\{(2,4,6,8),\,(1,2,7,8),\,(1,4,5,8),\,(5,6,7,8),\,(1,2,5,7),\,(3,4,5,6),\\
    &\,(2,3,6,7),\,(1,2,3,4)\}\,\,.
\end{align*}
Those in $I_-$ effectively swap the discrete torsion signs $\e_{f_{\gamma}}(\delta)$, while those in $I_+$ leave them alone. Interestingly, none of these 14 combinations change the signs in the $\a$, $\b$ or $\a\b$ sectors. Thus we have the set of dualities:
\begin{equation}
\begin{aligned}
    I_-:& M_{j,k} \rightarrow M_{4-j,k} \\
    I_+:& M_{j,k} \rightarrow M_{j,k} 
\end{aligned}\,\,.
\end{equation}
Both of these maps take type IIA string theory to type IIA and IIB to IIB. 
When $k=4$, these are the dualities found in \cite{Chuang:2004th}. However, for $k=0,1,2,3$ these are dualities between singular manifolds, which were not found in their analysis. Note there is no combination of T-dualities which change $k$.


\section{Spin(7) Mirror Maps for Connected Sums}\label{sect:oldspin7example}

In this section we consider mirror maps for Spin(7) manifolds realized as generalized connected sums (GCS) \cite{Braun:2018joh} 
and show that these agree with our results obtained above.

\subsection{Constructing Spin(7) Manifolds as Generalized Connected Sums}

As a preparation, let us briefly review the construction of GCS Spin(7) manifolds of \cite{Braun:2018joh}. The building blocks from which such Spin(7) are formed are a asymptotically cylindrical Calabi-Yau fourfold $Z_+$ with asymptotic neck region $X_3 \times S^1 \times I$ for a Calabi-Yau threefold $X_3$ and an interval $I$, and an asymptotically cylindrical $G_2$ manifold $Z_-$ with neck region $X_3 \times I$. Taking $Z_-\times S^1$ and identifying the isomorphic neck regions $X_3 \times S^1 \times I$, we may then form a compact eight-dimensional manifold $Z$ as the generalized connected sum 
\begin{equation}\label{eq:spin7GCS}
Z = Z_+  \,\#\, \left[ Z_- \times S^1\right] \, .
\end{equation}
Based on a number of observations, it has been conjectured in \cite{Braun:2018joh} that there exists a Ricci flat metric of holonomy Spin(7) on such manifolds. The evidence for this is as follows. First of all, the examples of Spin(7) manifolds realized as resolutions of $T^8/\Gamma$ for $\Gamma$ a finite subgroup of Spin(7) given in \cite{joyce1996spin7} allow precisely such a decomposition. We reviewed a decomposition such as \eqref{eq:spin7GCS} below in Section \ref{sect:spin7joyceasTCS}. Second, compactifications of heterotic string theory on TCS $G_2$ manifolds should have a lift to M-Theory on a Spin(7) manifold which can also be decomposed into two pieces. By applying an appropriate fibrewise duality map to a TCS $G_2$ manifold, the authors of 
\cite{Braun:2018joh} argued that one finds a decomposition such as \eqref{eq:spin7GCS} on the M-Theory side and checked the equivalence of the spectra of light fields in a few examples. Finally, acyl $G_2$ manifolds $Z_-$ can be realized as (a resolution of) a quotient $\left( X_3 \times \mathbb{R}\right) /\Z_2$ in which the $\Z_2$ acts as an anti-holomorphic quotient on $X_3$ and as $t \rightarrow -t $ on $\R$. In this case, $Z$ can be globally described as a resolution of an anti-holomorphic quotient of a suitably chosen Calabi-Yau fourfold $Y$, recovering the construction of \cite{joyce1996spin7_new}. 

For an acyl Calabi-Yau fourfold $Z_+$ and an acyl $G_2$ manifold $Z_-$ given as a resolution of $\left( X_3 \times \mathbb{R}\right) /\Z_2$, the Betti numbers of $Z$ are found to be \cite{Braun:2018joh}
\begin{equation}\label{eq:bettisgcsspin7}
\begin{aligned}
b^1(Z) & = 0 \\  
b^2(Z) & = n_+^2 + n_-^2 + b^2_e \\  
b^3(Z) & = n_-^2 + n_-^3 + n^3_+ \\  
b^4(Z) & = n_-^3 + n_-^4 + n_+^4 + b^2_o + b^3_o + b^3_e + b^4_e
\end{aligned}
\end{equation}
Here $n_\pm^i$ are the kernels of the restriction maps 
\begin{equation}
\begin{aligned}
\beta_+^i : &&H^i(Z_+,\Z) &\rightarrow H^i(X_3 \times \mathbb{S}^1,\Z)  \\
\beta_-^i : &&H^i(Z_-,\Z) &\rightarrow H^i(X_3,\Z)  \,.
\end{aligned}
\end{equation}
and $b^i_o$ and $b^i_e$ are the dimensions of the odd/even subspaces of the $i$-th cohomology group of $X_3$ under the action $\Z_2$. Furthermore, we have assumed that the images of $\beta_+^2$ and $\beta_+^4$ are surjective and that $H^3(Z_+) = \ker \beta^3_+$ holds \footnote{This last assumptions is slightly weaker than the assumptions made for technical simplicity in \cite{Braun:2018joh}. By following the same analysis presented there, it is straightforward to see that \eqref{eq:bettisgcsspin7} holds in the present case.}.

By using the fact that there exists a single covariantly constant spinor on $Z$, it follows that
\begin{equation}
b^4_-(Z) +1 = -8 + \tfrac13 \left(2-b^2(Z)+b^3(Z)+b^4(Z) \right) \, ,
\end{equation}
and we can compute 
\begin{equation}\label{eq:gluedspin7modulidimension}
b^2(Z) + b^4_-(Z) + 1 = \tfrac23 - 8 + \tfrac13\left(n^4_+ + 2 n^2_+ + n^3_+ \right) + \tfrac13\left(3n^2_- + 3 n^3_- + 2 (b^2_o + b^2_e) + 2 b^3_o  \right) \, .
\end{equation}
Here, we have used that for anti-holomorphic involutions $b^3_o = b^3_e$ and $b^4_e = b^2_o$ holds.

\subsection{A Mirror Map for GCS Spin(7) Manifolds}\label{sect:gcsmirrormap}

As shown in \cite{Shatashvili:1994zw}, exactly marginal deformations of Spin(7) sigma models are counted by \eqref{eq:gluedspin7modulidimension}, so
a mirror map for a Spin(7) manifold $Z$ must produce another manifold $Z^\vee$ such that
\begin{equation}\label{eq:spin7mirrorcondition}
b^2(Z) + b^4_-(Z) + 1 = b^2(Z^\vee) + b^4_-(Z^\vee) + 1 \, .
\end{equation}
Furthermore, such a map can be the result of an application of four T-dualities along a calibrated $T^4$ fibration \cite{Acharya:1997rh} \footnote{The moduli space of a Cayley (calibrated) four-cycle $N$ inside a Spin(7) manifold has dimension $-\tfrac12 N \cdot N$, so that we can at best hope to approximate such a fibration in a collapsed limit.}. The GCS decomposition \eqref{eq:spin7GCS} suggests how such a structure might be realized. The acyl Calabi-Yau fourfold $Z_+$ has a SYZ fibration by $T^4$ which becomes the $T^3$ SYZ fibre of $X_3$ times a circle $S^1$ in the neck region. On $Z_- \times S^1$, the circle simply becomes the product $S^1$ while the $T^3$ SYZ fibre of $X_3$ sits inside $Z_-= \left( X_3 \times \R \right)/\Z_2$. We hence expect to find a Spin(7) mirror by performing four T-dualities along this $T^4$.

This motivates the following construction: for a Spin(7) manifold realized as a GCS as in \eqref{eq:spin7GCS}, a mirror is given by\footnote{We would like to thank Michele del Zotto for suggesting this construction.}
\begin{equation}
 Z^\vee = Z_+^\vee  \,\#\, \left[ Z_-^\vee \times S^1\right] \, .
\end{equation}
which are glued along a neck region with is isomorphic to $X_3^\vee \times S^1 \times I$. In particular, $Z_-^\vee$ is constructed from an antiholomorphic involution of $X_3^\vee$ as $Z_-^\vee = \left(X_3^\vee \times \R\right)\Z_2$. 

In the following, we will collect some evidence for this proposal by showing that \eqref{eq:spin7mirrorcondition} indeed holds for this construction. In order to prove this, we will stick to the same simplifying assumptions under which \eqref{eq:bettisgcsspin7} holds. Our main task is to work out how the topology of $Z_\pm^\vee$ is related to that of $Z_\pm$. This can be done as follows. There is a compact Calabi-Yau fourfold $Y$ realized 
\begin{equation}\label{eq:decompcy4cy4}
Y = Z_+ \# Z_+ 
\end{equation}
realized by gluing two copies of $Z_+$ along $X_3 \times S^1 \times I$, and a $G_2$ manifold $M$ realized as
\begin{equation}
M = Z_- \# Z_-  = \left( X_3 \times S^1\right) /\Z_2
\end{equation}
by gluing two copies of $Z_-$ along $X_3 \times I$. For both of these compact geometries, 
there are mirror maps which act in the usual way, i.e.
\begin{equation}
\begin{aligned}
h^{1,1}(Y) &= h^{3,1}(Y^\vee)\\
h^{2,1}(Y) & = h^{2,1}(Y^\vee)\\
h^{3,1}(Y) &= h^{1,1}(Y^\vee)
\end{aligned}
\end{equation}
and 
\begin{equation}
b^2(M) + b^3(M) =  b^2(M^\vee) + b^3(M^\vee) \, . 
\end{equation}
Furthermore, $Y^\vee$ and $M^\vee$ now have the decompositions
\begin{equation}
Y^\vee = Z_+^\vee \# Z_+^\vee 
\end{equation}
glued along $X_3^\vee \times S^1 \times I$, and 
\begin{equation}
M^\vee = Z_-^\vee \# Z_-^\vee = \left( X_3^\vee \times S^1\right) /\Z_2\, .  
\end{equation}
glued along $X_3^\vee \times I$. Using these relations is the key to find the topology of $Z_\pm^\vee$ in terms of $Z_\pm$. 

Let us now work out the resulting relations in detail. Starting with $Y$, the Mayer-Vietoris sequence for the decomposition 
\eqref{eq:decompcy4cy4} gives
\begin{equation}
\begin{aligned}
b^2(Y) &= 2 n^2_+ + h^{1,1}(X_3) + 1 \\
b^3(Y) &= 2 n^3_+ \\
b^4(Y) &= 2 n^4_+ + 2 h^{1,1}(X_3) + 4 h^{2,1}(X_3) + 4
\end{aligned}\, .
\end{equation}
As $Y$ is a Calabi-Yau fourfold and $h^{2,1}(Y)=0$ there is the relation
\begin{equation}
h^{3,1}(Y) = \tfrac16 b^4(Y) - \tfrac23 b^2(Y) - \tfrac{23}{3}
\end{equation}
so that
\begin{equation}
h^{1,1}(Y^\vee) + h^{3,1}(Y^\vee) =  \tfrac13\left(n^4_+ +2n^2_+ \right) + \tfrac23\left(h^{1,1}(X_3) + h^{2,1}(X_3)\right) - \tfrac{23}{3}
\end{equation}
The mirror map acting on $Y$ must leave the above expression invariant. As this mirror map also maps $X_3$ to $X_3^\vee$, so that $h^{1,1}(X_3) + h^{2,1}(X_3)$ is preserved, it follows that $n^4_+ +2n^2_+$ must also be invariant under the mirror map acting on $Z_+$. Furthermore, $h^{2,1}(Y) = h^{2,1}(Y^\vee)$ implies that $n^3_+$ is the same for $Z_+$ and $Z_+^\vee$. 

Let us now discuss $M$. Here, the Mayer-Vietoris sequence yields
\begin{equation}
\begin{aligned}
b^2(M) &= b^2_e + 2 n^2_- \\ 
b^3(M) &= b^2_o + b^3_e + 2 n^3_- 
\end{aligned}\, . 
\end{equation}
Under the mirror map acting on the $G_2$ manifold $M$, $b^2(M) + b^3(M) = b^2(M^\vee) + b^3(M^\vee)$. As $b^2_e + b^2_o + b^3_e = 
h^{1,1}(X_3) + h^{2,1}(X_3) + 1$ for anti-holomorphic involutions, this expression is preserved by the mirror map. It hence follows that $n^2_- + n^3_-$ must also be left invariant under the mirror map acting on $Z_-$. 

Altogether, we have shown that the expressions $n^4_+ +2n^2_+$, $n^2_- + n^3_-$ and $b^2_e + b^2_o + b^3_e$ are all left invariant under an application of the mirror map acting on $Z_+$, $Z_-$ and $X_3$. It then follows that the expression \eqref{eq:gluedspin7modulidimension} for GCS Spin(7) manifolds is preserved under the mirror map, i.e. \eqref{eq:spin7mirrorcondition} holds. 

Note that we have not provided an explicit construction of mirrors for $Z_+$ and $Z_-$, but only used the topological constraints they have to satisfy to arrive at this conclusion. It should be possible to give a construction of acyl Calabi-Yau fourfolds from projecting five-dimensional tops as has been done for acyl Calabi-Yau threefolds in \cite{Braun:2016igl}. This would in turn allow to derive combinatorial formulae for the topological invariants of $Z_+$ and $Z_+^\vee$, which in turn must imply that $n^4_+ +2n^2_+$ does not change under the mirror map. 

It is of course straightforward to describe mirrors of $X_3$, but the definition of $Z_-$ furthermore involves specifying an antiholomorphic action of $\Z_2$ on $X_3$ and a resolution of the orbifold singularities of $\left(X_3 \times \R \right) / \Z_2$. Clearly, $b^2_e+b^2_o + b^3_e = h^{1,1}(X_3) + h^{2,1}(X_3) + 1$ does not depend on the details of the antiholomorphic involution chosen. Furthermore, in case there exists a resolution of $\left(X_3 \times S^1 \right) / \Z_2$ we have \cite{2017arXiv170709325J}
\begin{equation}
b^i(M) = b^i\left(\left(X_3\times S^1\right)/\Z_2\right) + b^{i-2}(L,\zeta) 
\end{equation}
where $L$ is the (real three-dimensional or empty) fixed locus of the involution and $\zeta$ is a possible twist. This potentially constrains which antiholomorphic involutions and which resolution can be chosen to construct $M^\vee$ and hence $Z_-^\vee$. 

As shown in \cite{Braun:2018joh}, the GCS construction of Spin(7) manifolds is closely related to the work of \cite{joyce1996spin7_new}, in which Spin(7) manifolds are found by resolving anti-holomorphic quotients of Calabi-Yau fourfolds. This offers another possible perspective on mirror maps of Spin(7) manifolds in general, and the ones considered here in particular.

\subsection{Examples}\label{sect:spin7joyceasTCS}

In this section we revisit the two examples of Spin(7) manifolds studies in Section \ref{sect:newspin7example} and show that the mirror map found there agrees with the GCS mirror map described above in Section \ref{sect:gcsmirrormap}.

\subsubsection{Example 1}

Let us first study the example of Section \ref{sect:spin7example1}, which has $c=(1,1,1,1)$ and $d=(0,1,1,1)$, and start by describing its GCS decomposition. Such a decomposition can be found by cutting the orbifold along $X_7 = \tfrac18$. At $X_7 = \tfrac18$, only two generators $\alpha$ and $\gamma$ act non-trivially on $X_1 \cdots X_6$, so that we can identify the neck region as $\tilde{X}_3 \times S^1 \times I$, where $X_7$ is a coordinate on $I$ and $X_8$ a coordinate on the $S^1$. Resolving the orbifold $\tilde{X}_3 = T^6 / \langle \alpha,\gamma\rangle$ produces the Calabi-Yau threefold $X_3$ with
\begin{equation}
\begin{aligned}
h^{1,1}(X_3) &= 19 \\  
h^{2,1}(X_3) &= 19 
\end{aligned}\,. 
\end{equation}

Restricting $X_7 \leq \tfrac18$, we find an acyl Calabi-Yau fourfold $\tilde{Z}_+ = T^7 \times \R / \langle \alpha,\beta,\gamma\rangle$. Two copies of $\tilde{Z}_+$ can be glued to form an Calabi-Yau orbifold $\tilde{Y}$, which has already been studied in \cite{Gopakumar:1996mu}. In their terminology, this case is the fourfold `model B', in which $\sigma$ is the Nikulin involution with invariants $(r,a,\delta) =(10,8,0)$. It can be described as $\left(K3 \times K3\right)/\Z_2$ with the $\Z_2$ acting as the Nikulin involution with invariants $(r,a,\delta) =(10,8,0)$ simultaneously on both K3 surfaces. The topology of the resolution $Y$ of $\tilde{Y}$ is given by
\begin{equation}\label{eq:ex1cy4a}
\begin{aligned}
h^{1,1}(Y) & = 24 \\
h^{2,1}(Y) & = 8  \\
h^{3,1}(Y) & = 24
\end{aligned}
\end{equation}
and $\chi(Y) = 288$. The relevant data of the decomposition $Y = Z_+ \# Z_+$ is 
\begin{equation}\label{eq:ex1cy4b}
\begin{aligned}
 n^2_+ &= 2 \\
n^3_+ &= 8 \\
n^4_+ &= 76
\end{aligned}
\end{equation}

Restricting $X_7 \geq \tfrac18$, we find the product of an $S^1$ and an acyl $G_2$ manifold $\tilde{Z}_- = T^6 \times \R / \langle \alpha,\gamma,\delta\rangle$, with $X_8$ being a coordinate on the product $S_1$. Two copies of this orbifold can be glued to the compact $G_2$ orbifold $\tilde{M}$ which has a unique resolution to a $G_2$ manifold $M$ with (see \cite{joyce1996II}) 
\begin{equation}
\begin{aligned}
b^2(M) &= 12\\ 
b^3(M) &= 43 \\ 
\end{aligned}\, .
\end{equation}
Furthermore, the action of $\delta$ at $X_7 = \tfrac14$ on $X_3$ is such that $b^2_e = 8$ and $b^2_o = 11$. This determines that
\begin{equation}
\begin{aligned}
n^2_- & = 2 \\
n^3_- & = 6 \\
n^4_- & = 6 
\end{aligned}
\end{equation}

As a check, one can now use \eqref{eq:bettisgcsspin7} to recover the Betti numbers of $Z$ given in \eqref{eq:bettispin7ex1} (note that the complete resolution corresponds to setting $k=4$ in \eqref{eq:bettispin7ex1}). As we have seen from the CFT analysis this Spin(7) manifold should be considered self-mirror. The same conclusion is reached by applying the GCS mirror map: both $Y$ and $M$ are self-mirror, so that 
\begin{equation}
Z_\pm = Z_\pm^\vee 
\end{equation}
and our Spin(7) mirror map gives $Z^\vee = Z$.

\subsubsection{Example 2}

Let us now study the example of Section \ref{sect:spin7example2}, which has $c=(1,0,1,0)$ and $d=(0,1,1,1)$. We can proceed in the same way as for the first example and cut along $X_7 = \tfrac18$. The neck region is again formed as $\tilde{X}_3 = T^6 / \langle \alpha,\gamma\rangle$ with 
\begin{equation}
\begin{aligned}
h^{1,1}(X_3) &= 19 \\  
h^{2,1}(X_3) &= 19 
\end{aligned}
\end{equation}

The acyl Calabi-Yau fourfolds $\tilde{Z}_+$ and its resolution $Z_+$ found by setting $x_7 \leq \tfrac18$ are the same as in the first example, so that we already know their topological data, \eqref{eq:ex1cy4a} and \eqref{eq:ex1cy4b}.  

The acyl $G_2$ orbifolds $\tilde{Z}_-$ and $\tilde{M}$ are different in this example, but $\tilde{M}$ is again one of the elementary examples of \cite{joyce1996II}. Its resolution is not unique and produces nine distinct $G_2$ manifolds $M_l$ with Betti numbers
\begin{equation}
\begin{aligned}
b^2(M_n) &= 8+ l \\ 
b^3(M_n) &= 47-  l \\ 
\end{aligned}\, ,
\end{equation}
for $l=0..8$. The action of $\delta$ at $X_7 = \tfrac14$ on $X_3$ is again such that $b^2_e = 8$ and $b^2_o = 11$. We now find that $l$ must be even and that
\begin{equation}\label{eq:nminusex2}
\begin{aligned}
n^2_- &= l/2 \\
n^3_-  &= 8-l/2 \\
n^4_-  &= 8-l/2 
\end{aligned}\, .
\end{equation}
This data again reproduces \eqref{eq:bettispin7ex1} from \eqref{eq:bettisgcsspin7} setting $l=2j$ and $k=4$ (again, only the case $k=4$ corresponds to a complete resolution). 

We are now ready to discuss the GCS mirror map for $Z$. We have $Z_+ = Z_+^\vee$ as before and $M_l^\vee = M_{8-l}$. This means that the GCS mirror map replaces $l \rightarrow 8-l$ in \eqref{eq:nminusex2}, so that it reproduces the CFT results $Z_j^\vee = Z_{4-j}$.  


\section*{Acknowledgements}

We would like to thank Michele del Zotto, Marc-Antoine Fiset, Sakura Sch\"afer-Nameki and Ashoke Sen for discussions and inspiration related to this project. The work of APB is supported by the ERC Consolidator Grant 682608 ``Higgs bundles: Supersymmetric Gauge Theories and Geometry'' (HIGGSBNDL). The research of AO was supported by the Perimeter Institute for Theoretical Physics. Research at Perimeter Institute is supported by the Government of Canada through the Department of Innovation, Science and Economic Development Canada and by the Province of Ontario through the Ministry of Economic Development, Job Creation and Trade. AO also acknowledges the support of NSERC through the PGS D scholarship.  


\appendix

\section{Discrete Torsion and Modular Invariance}\label{sect:introdtandmodinv}

To set the stage and outline our strategy, let us review a few basic facts about (generalized) discrete torsion for strings on orbifolds following \cite{Gaberdiel:2004vx}. Crucially, the definition of string theory on orbifolds in general involves an assignment of discrete torsion phases \cite{vafa1986modular}. String theory on a orbifold of $T^n$ by a group $\Gamma$ is built from the untwisted sector $\mathcal{H}_e$ composed of $\Gamma$-invariant states, as well as a twisted sector $\mathcal{H}_g$ for every non-trivial group element $g$ of $\Gamma$. To find the states in the twisted sectors $\mathcal{H}_g$, we need to study the action of other group elements $h\neq g$ on $\mathcal{H}_g$. This action in general involves the assignment of phases, 
\begin{equation}
h|_{\mathcal{H}_g} = \epsilon_g(h) h^0 |_{\mathcal{H}_g}\, ,
\end{equation}
where $h^0$ refers to the usual action of h in the g-twisted sector as expected from the orbifold group action on the coordinates. These discrete torsion phases must form a representation of $\Gamma$ and furthermore must satisfy \cite{vafa1986modular}
\begin{equation}\label{eq:normaldtmodconstraint}
\epsilon_g(h) = \epsilon_{h^cg^d}(h^a g^b) \hspace{.5cm} \mbox{for} \hspace{.3cm}  ad-bc = 1
\end{equation}
to guarantee modular invariance. As the twisted sector associated with a group element $g$ typically decomposes as
\begin{equation}\label{eq:tsdecomp}
\mathcal{H}_g = \oplus_f \mathcal{H}_{g,f}\, ,
\end{equation}
e.g. in case $g$ has several fixed points labelled by $f_g$, a different assignment of phases $\epsilon_{f_g}(h)$ for each $f_g$ is possible \cite{Gaberdiel:2004vx}. Of course, these still have to form a representation of $\Gamma$. 

If we choose to include such `generalized' discrete torsion phases $\epsilon_{f_g}(h)$ in our model, modular invariance must be reconsidered. The partition function for our models can be written as
\begin{equation}\label{partitioncomponentdef}
Z(q,\bar{q}) = \frac{1}{|\Gamma|} \sum_{h,g \in \Gamma} \Tr_{\mathcal{H}_h} \left(g q^{L_0 -c/24}\bar{q}^{L_0 -c/24} \right) \equiv \frac{1}{|\Gamma|} \sum_{h,g \in \Gamma} Z_{h;g} \, .
\end{equation}
where $Z_{h;g}$ refers to the partition function component restricted to the h twisted sector, as in the summation in the middle. Modular invariance then implies
\begin{equation}\label{Tconstraintpartition}
Z(\tau+1) = Z(\tau) \,\,\rightarrow \,\, Z_{g;e}(\tau + 1) = Z_{g;g}
\end{equation}
from which $\epsilon_{f_g}(g)=1$ follows, and
\begin{equation}\label{Sconstraintpartition}
Z(-1/\tau) = Z(\tau) \,\,\rightarrow \,\,  Z_{g;h}(-1/\tau) = Z_{h;g}(\tau)
\end{equation}
which constrains possible assignments of the $\epsilon_{f_g}(h)$ by linking them to the phases $\epsilon_{f_h}(g)$. 

Although modular invariance for bosonic strings at one loop is sufficient to guarantee modular invariance at higher genus if $\epsilon_{f_g}(h) = \epsilon_{f'_g}(h)$ for all $f_g,f_g'$, this is not the case for more general assignments. However, studying solutions to the above constraints at least provide us with necessary conditions, which will be enough for our purposes. Furthermore, we are only going to study partition functions of bosonic strings. Although it is generally believed that modular invariance of the bosonic string partition function is necessary and sufficient for modular invariance of the full superstring theory, higher genus amplitudes again present a caveat to this analysis, see \cite{Gaberdiel:2004vx} for a more detailed discussion. For the examples we are presenting, these subtleties are alleviated by the fact that we can match them to known smooth geometries obtained by a smoothing of the orbifolds in question. 

For the examples discussed in this paper, the computation of partition functions is significantly simplified by the fact that all of the elements of the orbifold group act diagonally on $T^7 = (S^1)^7$ or $T^8= (S^1)^8$. For the sake of brevity, we have omitted the details of these computations. 


\section{Discrete Torsion Analysis for the $G_2$ orbifold}

In this appendix, we derive the necessary conditions on discrete torsion phases for the $G_2$ model introduced in section \ref{G2}. We work out the representation matrices for the orbifold elements in the highest weight states of the different twisted sectors. These matrices will have discrete torsion signs showing up, which are then constrained by trace relations coming from the S-transformation. 

The orbifold we are interested in is defined by
\begin{center}
\begin{tabular}{ |c|c|c|c|c|c|c|c| } 
 \hline
  & $X^1$ & $X^2$ & $X^3$ & $X^4$ & $X^5$ & $X^6$ & $X^7$\\ 
  \hline
 $\alpha$ & + & + & + & - & - & - & -\\ 
  $\beta$ & + & - & - & + & + & -$\frac{1}{2}$ & - \\ 
  $\gamma$ & - & + & - & + & - & + & -$\frac{1}{2}$\\
  $\sigma_2$ & + & +$\frac{1}{2}$ & + & +$\frac{1}{2}$ & + & + & + \\ 
 \hline
\end{tabular}
\end{center}

As discussed in the section 2, we do not need to analyse all the different twisted sectors in this orbifold. Instead, we would focus on the particular sectors twisted under the action of $\alpha$, $\beta$, $\gamma$, $\alpha\beta$, $\sigma_2$ and $\sigma_2\alpha\beta$. The $\alpha$, $\beta$, and $\gamma$ sectors are relevant because they are the only ones that contribute to the ground state spectrum of the orbifold string theory, and the others are needed only in order to fix the discrete torsion phases.

\subsubsection*{$\a$ sector}\label{G2aplha}

The $\alpha$-twisted sector can be decomposed into 16 smaller sectors corresponding to the fixed points of the action of $\alpha$. They can be labelled by the different values of $\{(X^4,X^5,X^6,X^7): X^i\in\{0,\frac{1}{2}\}\}$. Now each of these twisted sectors localised at the fixed points, have a highest weight state of zero momentum and zero winding. We want to find the representation matrices for the orbifold elements in the basis of these highest weight states. Under the free action of $\langle\beta,\gamma,\sigma_2\rangle$, we get two 8D irreducible representations corresponding to the two $X^5$ choices. 

We can then assign coordinate labels for the basis states $\ket{j}_{\alpha}^{f_{\alpha}}$, where index $f_{\alpha}=\{1,2\}$ corresponds to the two choices of $X^5$ and j=1,2,..,8 enumerates the different choices of the other 3 fixed-point coordinates ($X^4,X^6,X^7$): 
\begin{equation}
\begin{split}
&\ket{1}_{\alpha}^{f_{\alpha}}\sim(0,0,0);\hspace{5pt}\ket{2}_{\alpha}^{f_{\alpha}}\sim\left(0,0,\frac{1}{2}\right); \hspace{5pt} \ket{3}_{\alpha}^{f_{\alpha}}\sim\left(0,\frac{1}{2},0\right);\hspace{5pt}\ket{4}_{\alpha}^{f_{\alpha}}\sim\left(0,\frac{1}{2},\frac{1}{2}\right);\\
&\ket{5}_{\alpha}^{f_{\alpha}}\sim\left(\frac{1}{2},0,0\right);\hspace{5pt}\ket{6}_{\alpha}^{f_{\alpha}}\sim\left(\frac{1}{2},0,\frac{1}{2}\right); \hspace{5pt} \ket{7}_{\alpha}^{f_{\alpha}}\sim\left(\frac{1}{2},\frac{1}{2},0\right);\hspace{5pt}\ket{8}_{\alpha}^{f_{\alpha}}\sim\left(\frac{1}{2},\frac{1}{2},\frac{1}{2}\right);
\end{split}
\end{equation}
In this basis, the orbifold generators act as follows:
\begin{equation}
\begin{aligned}
\alpha|_{\mathcal{H}_{\alpha}^{f_{\alpha}}}&=\mathbb{I}_{8\times 8} \hspace{10pt} 
&\beta|_{\mathcal{H}_{\alpha}^{f_{\alpha}}}&=\begin{pmatrix}
\ket{1}_{\alpha}^{f_{\alpha}}\leftrightarrow\ket{3}_{\alpha}^{f_{\alpha}}\\
\ket{2}_{\alpha}^{f_{\alpha}}\leftrightarrow\ket{4_{\alpha}^{f_{\alpha}}}\\
\ket{5}_{\alpha}^{f_{\alpha}}\leftrightarrow\ket{7}_{\alpha}^{f_{\alpha}}\\
\ket{6}_{\alpha}^{f_{\alpha}}\leftrightarrow\ket{8}_{\alpha}^{f_{\alpha}}
\end{pmatrix}
\\
\gamma|_{\mathcal{H}_{\alpha}^{f_{\alpha}}}&=\begin{pmatrix}
\ket{1}_{\alpha}^{f_{\alpha}}\leftrightarrow\ket{2}_{\alpha}^{f_{\alpha}}\\
\ket{3}_{\alpha}^{f_{\alpha}}\leftrightarrow\ket{4}_{\alpha}^{f_{\alpha}}\\
\ket{5}_{\alpha}^{f_{\alpha}}\leftrightarrow\ket{6}_{\alpha}^{f_{\alpha}}\\
\ket{7}_{\alpha}^{f_{\alpha}}\leftrightarrow\ket{8}_{\alpha}^{f_{\alpha}}
\end{pmatrix}\hspace{10pt}
&\sigma_2|_{\mathcal{H}_{\alpha}^{f_{\alpha}}}&=\begin{pmatrix}
\ket{1}_{\alpha}^{f_{\alpha}}\leftrightarrow\ket{5}_{\alpha}^{f_{\alpha}}\\
\ket{2}_{\alpha}^{f_{\alpha}}\leftrightarrow\ket{6}_{\alpha}^{f_{\alpha}}\\
\ket{3}_{\alpha}^{f_{\alpha}}\leftrightarrow\ket{7}_{\alpha}^{f_{\alpha}}\\
\ket{4}_{\alpha}^{f_{\alpha}}\leftrightarrow\ket{8}_{\alpha}^{f_{\alpha}}
\end{pmatrix}\, ,\hspace{10pt}
\end{aligned}
\end{equation}
where $\mathcal{H}_{\alpha}=\bigoplus_{f_{\alpha}} \mathcal{H}_{\alpha}^{f_{\alpha}}$ and $\mathcal{H}_{\alpha}^{f_{\alpha}}$ is the space spanned by the highest weight states $\ket{j}_{\alpha}^{f_{\alpha}}$. The representation matrices for the generators after removing spurious phases are
\begin{equation}
\begin{aligned}
\alpha|_{\mathcal{H}_{\alpha}^{f_{\alpha}}}&=\mathbb{I}_{8\times8}
&\beta|_{\mathcal{H}_{\alpha}^{f_{\alpha}}}&=\begin{pmatrix}
0&\mathbb{I}_{2\times2}&0&0\\
\mathbb{I}_{2\times2}&0&0&0\\
0&0&0&\mathbb{I}_{2\times2}\\
0&0&\mathbb{I}_{2\times2}&0
\end{pmatrix}\\
\gamma|_{\mathcal{H}_{\alpha}^{f_{\alpha}}}&=\begin{pmatrix}
H&0&0&0\\
0&H&0&0\\
0&0&H&0\\
0&0&0&H
\end{pmatrix}
&\sigma_2|_{\mathcal{H}_{\alpha}^{f_{\alpha}}}&=\begin{pmatrix} 0&\mathbb{I}_{4\times4}\\ \mathbb{I}_{4\times4}&0\\ \end{pmatrix}
\end{aligned}
\end{equation}
where H=$\begin{pmatrix}0&1\\1&0\\ \end{pmatrix}$.

\subsubsection*{$\beta$ sector}\label{G2beta}

The $\beta$ sector only differs from the $\alpha$ case in the fixed point coordinate labels. Here they are given by two choices each for the set of coordinates ($X^2,X^3,X^6,X^7$). $\alpha$ behaves the same way as $\beta$ did in the $\alpha$-sector, $\gamma$ mixes the 2 $X^7$ choices as before, and $\sigma_2$ mixes the choices for $X^2$. So, no discrete torsion phase arises in this sector either.

\subsubsection*{$\gamma$ sector}\label{G2gamma}

Analogous to the $\alpha$ sector, there are 16 highest weight states in the $\gamma$ sector which can be identified by their fixed point coordinate labels ($X^1,X^3,X^5,X^7$). Under the free action of $\langle \alpha,\beta,\sigma_2\rangle$, we get eight 2D irreducible representations corresponding to the two choices for $X^1$, $X^3$, and $X^5$ each. 

Let us now assign coordinate labels for the basis states $\ket{j}_{\gamma}^{f_{\gamma}}$, where index $f_{\gamma}=\{1,2,...,8\}$ corresponds to the 8 choices of ($X^1,X^3,X^5$) and j=1,2 enumerates the two choices for $X^7$:
\begin{equation}
    \ket{1}_{\gamma}^{f_{\gamma}}\sim\left(X^7=\frac{1}{4}\right);\hspace{5pt} \ket{2}_{\gamma}^{f_{\gamma}}\sim\left(X^7=\frac{3}{4}\right)
\end{equation}

Then the action of the orbifold generators can be obtained from their action on the coordinate labels, as follows:

\begin{equation}
    \alpha|_{\mathcal{H}_{\gamma}^{f_{\gamma}}}=\left(\ket{1}_{\gamma}^{f_{\gamma}}\leftrightarrow\ket{2}_{\gamma}^{f_{\gamma}}\right)=\beta|_{\mathcal{H}_{\gamma}^{f_{\gamma}}};\hspace{5pt}\gamma|_{\mathcal{H}_{\gamma}^{f_{\gamma}}}=\text{id.}=\sigma_2|_{\mathcal{H}_{\gamma}^{f_{\gamma}}}
\end{equation}

Removing spurious phases by exploiting commutation relations of the representation matrices, we get:

\begin{equation}
\alpha|_{\mathcal{H}_{\gamma}^{f_{\gamma}}}=H; \hspace{10pt} \beta|_{\mathcal{H}_{\gamma}^{f_{\gamma}}}=\epsilon_{f_{\gamma}}(\alpha\beta)H; \hspace{10pt}\gamma|_{\mathcal{H}_{\gamma}^{f_{\gamma}}}=\mathbb{I}_{2\times2}; \hspace{10pt} \sigma_2|_{\mathcal{H}_{\gamma}^{f_{\gamma}}}=\begin{pmatrix}
\epsilon_1&0\\
0&\epsilon_2
\end{pmatrix}
\end{equation}
where $\epsilon_{f_{\gamma}}(\alpha\beta)=\pm 1$. The relation $\sigma_2\alpha=\alpha\sigma_2$ yields $\epsilon_1=\epsilon_2=\epsilon_{f_{\gamma}}(\sigma_2)=\pm1$:
\begin{equation}
\sigma_2=\epsilon_{f_{\gamma}}(\sigma_2)\mathbb{I}_{2\times2}
\end{equation}
So there are two choices of discrete torsion signs available in each irreducible representation. 

\subsubsection*{$\alpha\beta$ sector}

The action of $\alpha\beta$ on the coordinates is given by: 

\begin{center}
\begin{tabular}{ |c|c|c|c|c|c|c|c| } 
 \hline
  & $X^1$ & $X^2$ & $X^3$ & $X^4$ & $X^5$ & $X^6$ & $X^7$\\ 
  \hline
  $\alpha\beta$ & + & - & - & - & - & +$\tfrac12$ & + \\
 \hline
\end{tabular}
\end{center}

In the $\alpha\beta$ twisted sector, lowest energy states are labelled by the half-integer mode $n_6$ taking values in $\pm \frac{1}{2}$, and coordinate labels ($X^2,X^3,X^4,X^5$). Now the irreducible representations are 8D, and are spanned by the ($n_6,X^2,X^4$) coordinate labels. Let us assign the basis states $\ket{j}_{\alpha\beta}^{f_{\alpha\beta}}$, $f_{\alpha\beta}=1,2,..4$ labels the different irreducible representations corresponding to the choices for ($X^3,X^5$):
\begin{equation}
\begin{split}
           &\ket{1}_{\alpha\beta}^{f_{\alpha\beta}}\sim\left(\frac{1}{2},0,0\right);\hspace{5pt}\ket{2}_{\alpha\beta}^{f_{\alpha\beta}}\sim\left(-\frac{1}{2},0,0\right);\hspace{5pt}\ket{3}_{\alpha\beta}^{f_{\alpha\beta}}\sim\left(\frac{1}{2},0,\frac{1}{2}\right);\\
           &\ket{4}_{\alpha\beta}^{f_{\alpha\beta}}\sim\left(-\frac{1}{2},0,\frac{1}{2}\right)\hspace{5pt}\ket{5}_{\alpha\beta}^{f_{\alpha\beta}}\sim\left(\frac{1}{2},\frac{1}{2},0\right);\hspace{5pt}\ket{6}_{\alpha\beta}^{f_{\alpha\beta}}\sim\left(-\frac{1}{2},\frac{1}{2},0\right);\\
           &\ket{7}_{\alpha\beta}^{f_{\alpha\beta}}\sim\left(\frac{1}{2},\frac{1}{2},\frac{1}{2}\right);\hspace{5pt}\ket{8}_{\alpha\beta}^{f_{\alpha\beta}}\sim\left(-\frac{1}{2},\frac{1}{2},\frac{1}{2}\right)\\
\end{split}
\end{equation}
In this basis, we can write the representation matrices for the orbifold elements by looking at their action on the basis states just as we did before for the previous sectors:
\begin{equation}\label{absectorreps}
\begin{split}
    &\alpha|_{\mathcal{H}_{\alpha\beta}^{f_{\alpha\beta}}}=\beta|_{\mathcal{H}_{\alpha\beta}^{f_{\alpha\beta}}}=\begin{pmatrix}H&0&0&0\\0&H&0&0\\0&0&H&0\\0&0&0&H
    \end{pmatrix}\\
    &\gamma|_{\mathcal{H}_{\alpha\beta}^{f_{\alpha\beta}}}=\begin{pmatrix}
    \epsilon_{f_{\alpha\beta}}^1(\gamma)\mathbb{I}_{4\times4}&0\\
    0&\epsilon_{f_{\alpha\beta}}^2\mathbb{I}_{4\times4}(\gamma)
    \end{pmatrix};\hspace{5pt} \sigma_2|_{\mathcal{H}_{\alpha\beta}^{f_{\alpha\beta}}}=\begin{pmatrix}
    0&0&0&\mathbb{I}_{2\times 2}\\0&0&\mathbb{I}_{2\times 2}&0\\0&\mathbb{I}_{2\times 2}&0&0\\\mathbb{I}_{2\times 2}&0&0&0
    \end{pmatrix} 
\end{split}
\end{equation}
Here, the discrete torsion sign shows up in the $\gamma$ matrix: $\epsilon_{f_{\alpha\beta}}^i(\gamma)=\pm 1$, i=1,2; as is expected from our $\gamma$ sector analysis and the S-transform relations. 

\subsubsection*{$\sigma_2$ sector}

The action of $\sigma_2$ on the coordinates is given by: 

\begin{center}
\begin{tabular}{ |c|c|c|c|c|c|c|c| } 
 \hline
  & $X^1$ & $X^2$ & $X^3$ & $X^4$ & $X^5$ & $X^6$ & $X^7$\\ 
  \hline
  $\sigma_2$ & + & +$\frac{1}{2}$ & + & +$\frac{1}{2}$ & + & + & + \\
 \hline
\end{tabular}
\end{center}

In the $\sigma_2$-twisted sector, the lowest energy states are labelled by the two half-integer winding numbers ($n_2$,$n_4$) each taking values in $\pm \frac{1}{2}$. We can then have the following assignment of basis states $\ket{j}_{\sigma_2}$, where j=1,2,3,4 :  

\begin{equation}
    \ket{1}_{\sigma_2}\sim\left(\frac{1}{2},\frac{1}{2}\right);\hspace{5pt}\ket{2}_{\sigma_2}\sim\left(\frac{1}{2},-\frac{1}{2}\right);\hspace{5pt}\ket{3}_{\sigma_2}\sim\left(-\frac{1}{2},\frac{1}{2}\right);\hspace{5pt}\ket{4}_{\sigma_2}\sim\left(-\frac{1}{2},-\frac{1}{2}\right)
\end{equation}
Looking at the action of the different generators on the basis states as listed above, we get the following 4D representation matrices:

\begin{equation}\label{sigma2irrrep}
    \alpha=\begin{pmatrix}
H&0\\
0&H
\end{pmatrix};\hspace{5pt}\beta=\begin{pmatrix}
0&0&1&0\\
0&0&0&1\\
1&0&0&0\\
0&1&0&0
\end{pmatrix};\hspace{5pt}\gamma=\epsilon_{\sigma_2}(\gamma)\mathbb{I}_{4\times4};\hspace{5pt}\sigma_2=\mathbb{I}_{4\times4}
\end{equation}
Note that the only non-trivial trace involving a discrete torsion sign in this sector is for $\gamma$.

\subsubsection*{$\sigma_2\alpha\beta$ sector}\label{G2sab}

The action of $\sigma_2\alpha\beta$ on the coordinates is given by: 

\begin{center}
\begin{tabular}{ |c|c|c|c|c|c|c|c| } 
 \hline
  & $X^1$ & $X^2$ & $X^3$ & $X^4$ & $X^5$ & $X^6$ & $X^7$\\ 
  \hline
  $\sigma_2\alpha\beta$ & + & -$\frac{1}{2}$ & - & -$\frac{1}{2}$ & - & +$\frac{1}{2}$ & + \\
 \hline
\end{tabular}
\end{center}
Just as the case with the $\alpha\beta$ sector, the lowest energy states are labelled by ($X^2,X^3,X^4,X^5,n_6$), each of which takes two values. The irreducible representation, as deduced from the action of the orbifold generators, corresponds to the labels ($n^6,X^2,X^4$), and can be organised in the basis $\ket{j}_{\sigma_2\alpha\beta}^{f_{\sigma_2\alpha\beta}}$, where j=1,2,..,8, and $f_{\sigma_2\alpha\beta}=1,2,..,4$:

\begin{equation}
\begin{split}
           &\ket{1}_{\sigma_2\alpha\beta}^{f_{\sigma_2\alpha\beta}}\sim\left(\frac{1}{2},\frac{1}{4},\frac{1}{4}\right);\hspace{5pt}\ket{2}_{\sigma_2\alpha\beta}^{f_{\sigma_2\alpha\beta}}\sim\left(-\frac{1}{2},\frac{1}{4},\frac{1}{4}\right);\hspace{5pt}\ket{3}_{\sigma_2\alpha\beta}^{f_{\sigma_2\alpha\beta}}\sim\left(\frac{1}{2},\frac{1}{4},\frac{3}{4}\right);\\
           &\ket{4}_{\sigma_2\alpha\beta}^{f_{\sigma_2\alpha\beta}}\sim\left(-\frac{1}{2},\frac{1}{4},\frac{3}{4}\right)\hspace{5pt}\ket{5}_{\sigma_2\alpha\beta}^{f_{\sigma_2\alpha\beta}}\sim\left(\frac{1}{2},\frac{3}{4},\frac{1}{4}\right);\hspace{5pt}\ket{6}_{\sigma_2\alpha\beta}^{f_{\sigma_2\alpha\beta}}\sim\left(-\frac{1}{2},\frac{3}{4},\frac{1}{4}\right);\\
           &\ket{7}_{\sigma_2\alpha\beta}^{f_{\sigma_2\alpha\beta}}\sim\left(\frac{1}{2},\frac{3}{4},\frac{3}{4}\right);\hspace{5pt}\ket{8}_{\sigma_2\alpha\beta}^{f_{\sigma_2\alpha\beta}}\sim\left(-\frac{1}{2},\frac{3}{4},\frac{3}{4}\right)\\
\end{split}
\end{equation}

After absorption of spurious phases via commutation relations and normalization of states, we get a discrete torsion sign arising in $\gamma$ as follows:
\begin{equation}
\begin{split}
    &\alpha|_{\mathcal{H}_{\sigma_2\alpha\beta}^{f_{\sigma_2\alpha\beta}}}=\begin{pmatrix}0&H&0&0\\H&0&0&0\\0&0&0&H\\0&0&H&0
    \end{pmatrix};\hspace{5pt}\beta|_{\mathcal{H}_{\sigma_2\alpha\beta}^{f_{\sigma_2\alpha\beta}}}=\begin{pmatrix}0&0&H&0\\0&0&0&H\\H&0&0&0\\0&H&0&0
    \end{pmatrix}\\
    &\gamma|_{\mathcal{H}_{\sigma_2\alpha\beta}^{f_{\sigma_2\alpha\beta}}}=\epsilon_{f_{\sigma_2\alpha\beta}}(\gamma)\mathbb{I}_{8\times 8};\hspace{5pt} \sigma_2|_{\mathcal{H}_{\sigma_2\alpha\beta}^{f_{\sigma_2\alpha\beta}}}=\begin{pmatrix}
    0&0&0&\mathbb{I}_{2\times 2}\\0&0&\mathbb{I}_{2\times 2}&0\\0&\mathbb{I}_{2\times 2}&0&0\\\mathbb{I}_{2\times 2}&0&0&0
    \end{pmatrix}\\
\end{split}
\end{equation}


\section{Discrete Torsion Analysis for the $Spin(7)$ Orbifolds}\label{app:spin7}

In this appendix, we will explicitly determine the allowed discrete torsion phases and constraints for the $Spin(7)$ orbifold in section 3 - the general structure follows the same logic as in appendix B.

First, let us recall the definition of the orbifold we are interested in - we focus on example 2 from section 3. This is a $T^8/ \mathbb{Z}_2^4$ orbifold where the group generators $\alpha$, $\beta$, $\gamma$ and $\delta$ act as:

\begin{center}
\begin{tabular}{ |c|c|c|c|c|c|c|c|c| } 
 \hline
  & $X^1$ & $X^2$ & $X^3$ & $X^4$ & $X^5$ & $X^6$ & $X^7$ & $X^8$\\ 
  \hline
  $\alpha$ & - & - & - & - & + & + & + & +\\ 
  $\beta$ & + & + & + & + & - & - & - & - \\ 
  $\gamma$ & $-\frac{1}2$ & - & + & + & $-\frac{1}2$ & - & + & +\\
  $\delta$ & - & + & $-\frac{1}2$ & + & $-\frac{1}2$ & + & $-\frac{1}2$ & + \\ 
 \hline
\end{tabular}
\end{center}
Let us now determine the allowed discrete torsion phases.


\subsubsection*{$\a$, $\beta$ and $\alpha\beta$ sectors}

We begin with $\a$. Here we have 16 twist fields in one-to-one correspondence with the fixed points of the action of $\a$, as outlined in section 3. These states are labelled by the choice of a coordinate set $\{(X^1,X^2,X^3,X^4)\,:\,X^i\in\{0,\,\frac1{2} \}\}$. Under the action of the rest of the group, in particular by $\langle \gamma, \delta\rangle$, these get grouped into 4 sets of 4, corresponding to 4 irreducible representations of the orbifold group. Each of these representations come with a highest weight state of zero momentum and winding, and our goal is to find the matrix representations of the group element in the basis of such highest weight states.

Explicitly, in this sector $\gamma$ and $\delta$ permute the $X^1$ and $X^3$ fixed points, and so we can label each of the 4 representations by a number $f_{\a}\in \{1,..,4\}$ corresponding to one of the 4 choices of $X^2$ and $X^4$. In other words, the 16 dimensional space of these highest weight states decomposes further into a sum of 4 dimensional spaces as $\mathcal{H}_\a=\oplus_{f_\a}\mathcal{H}_\a^{f_\a}$. In each $\mathcal{H}_\a^{f_\a}$, the basis of states then consists of vectors $\ket{i}^{f_{\a}}_{\a}$, where the label $i\in\{1,..,4\}$ represents one of the 2-tuples in $\{(X^1,X^3)\,:\,X^1,X^3\in\{0,\frac1{2}\}\}$. Explicitly, we set:
\begin{equation}
\begin{split}
    &\ket{1}^{f_{\a}}_{\a}=(0,0)\,,\quad \ket{2}^{f_{\a}}_{\a}=(1/2,0)\,\,,\\
    &\ket{3}^{f_{\a}}_{\a}=(0,1/2)\,,\quad \ket{4}^{f_{\a}}_{\a}=(1/2,1/2)\,\,.
\end{split}
\end{equation}
In this basis, $\a$ acts trivially, $\beta$ acts diagonally and the action of $\gamma$ and $\delta$ is:
\begin{gather}
    \gamma|_{\mathcal{H}_\a^{f_\a}}\,:\,\begin{pmatrix}\ket{1}^{f_{\a}}_{\a}\leftrightarrow\ket{2}^{f_{\a}}_{\a} \\ \ket{3}^{f_{\a}}_{\a}\leftrightarrow\ket{4}^{f_{\a}}_{\a}\end{pmatrix}\,,\quad\hbox{and}\quad \delta|_{\mathcal{H}_\a^{f_\a}}\,:\,\begin{pmatrix}\ket{1}^{f_{\a}}_{\a}\leftrightarrow\ket{3}^{f_{\a}}_{\a} \\ \ket{2}^{f_{\a}}_{\a}\leftrightarrow\ket{4}^{f_{\a}}_{\a}\end{pmatrix}\,\,.
\end{gather}
Introducing discrete torsion phases, by an appropriate choice of normalization of the basis vectors the matrix representations of $\gamma$, $\delta$ and $\beta$ take the form:
\begin{equation}
    \gamma|_{\mathcal{H}_\a^{f_\a}}=\begin{pmatrix}H & 0 \\ 0 & H\end{pmatrix}\,,\quad
    \delta|_{\mathcal{H}_\a^{f_\a}}=\begin{pmatrix}0 & 0 & 1 & 0 \\ 0 & 0 & 0 & e^{i\theta} \\ 1 & 0 & 0 & 0 \\ 0 & e^{-i\theta} & 0 & 0\end{pmatrix}\,\quad\hbox{and}\quad
    \beta|_{\mathcal{H}_\a^{f_\a}}=\begin{pmatrix}\e_1 & 0 & 0 & 0\\ 0 & \e_2 & 0 & 0 \\ 0 & 0 & \e_3 & 0 \\ 0 & 0 & 0 & \e_4\end{pmatrix}\,\,.
\end{equation}
with $\e_i^2=1$. The requirement that all group elements commute sets the phases in $\delta|_{\mathcal{H}_\a^{f_\a}}$ to 1, and forces all $\e_i$ to be equal, $\e_i\equiv\e_{f_\a}(\beta)$ for all $i$, so that $\beta=\e_{f_\a}(\beta)\cdot \mathbb{I}_{4\times 4}$. So, the 16 signs we would expect in $Z_{\alpha;\beta}$ are identified in 4's, reducing to only 4 degrees of freedom $\e_{f_\a}(\beta)$. 

The $\beta$ analysis is virtually identical, only differing by the coordinate labelling of states (e.g. the fixed points now correspond to the set $\{(X^5,X^6,X^7,X^8)\,:\,X^i\in\{0,\,\frac1{2} \}\}$) and the exchanging of the roles of $\a$ and $\beta$. Once again, the 16 signs get identified in 4's, and so we end up with the 4 sign degrees of freedom $\e_{f_\beta}(\alpha)$ for $f_\beta=1..4$.

Next, let's do $\a\beta$. In this case $\a\beta$ sends $X^i\rightarrow -X^i$ for all $i$, and so we have 256 fixed points corresponding to the choices $X^i\in\{0,\frac1{2}\}$. $\a$ and $\beta$ clearly act diagonally, $\gamma$ permutes $X^1$ and $X^5$, and $\delta$ permutes $X^3$, $X^5$ and $X^7$. We can choose two of these as labels for our states, and in particular choose $X^1$ and $X^3$. This groups our states into 64 sets of 4 - i.e. we have a decomposition $\mathcal{H}_{\a\beta}=\oplus_{f_{\a\beta}=1}^{64}\mathcal{H}_{\a\beta}^{f_{\a\beta}}$, with each $\mathcal{H}_{\a\beta}^{f_{\a\beta}}$ 4 dimensional. It is then easy to see that the actions of $\gamma|_{\mathcal{H}_{\a\beta}^{f_{\a\beta}}}$ and $\delta|_{\mathcal{H}_{\a\beta}^{f_{\a\beta}}}$ are the same as they were in e.g. the $\a$ sector, and so they take the same form (also without discrete torsion phases). For $\a$ and $\beta$, we may set them both equal (in order to impose $\a\beta|_{\mathcal{H}_{\a\beta}^{f_{\a\beta}}}=\mathbb{I}_{4\times 4}$) to:
\begin{equation}
    \a|_{\mathcal{H}_{\a\beta}^{f_{\a\beta}}}=\beta|_{\mathcal{H}_{\a\beta}^{f_{\a\beta}}}=\begin{pmatrix}\e_1 & 0 & 0 & 0\\ 0 & \e_2 & 0 & 0 \\ 0 & 0 & \e_3 & 0 \\ 0 & 0 & 0 & \e_4\end{pmatrix}\,\,,
\end{equation}
so that $\a\beta=1$ when $\e_i^2=1$. The commutation constraints set all $\e_i$ equal, and we call them $\e_{f_{\a\beta}}(\a,\,\beta)$ so that $\a|_{\mathcal{H}_{\a\beta}^{f_{\a\beta}}}=\beta|_{\mathcal{H}_{\a\beta}^{f_{\a\beta}}}=\e_{f_{\a\beta}}(\a,\,\beta)\cdot \mathbb{I}_{4\times 4}$ (we use the notation $\e_{f_{\a\beta}}(\a,\,\beta)$ to denote the fact that both $\alpha$ \textit{and} $\beta$ have discrete torsion signs in the $\alpha\beta$ sector), and we end up with 64 sign degrees of freedom $\e_{f_{\a\beta}}(\a,\beta)$. 


\subsubsection*{$\delta$ sector}

Now let's do the $\delta$ sector. Here we get 16 fixed points in $(X^1,X^3,X^5,X^7)$ grouped into 2 sets of 8 by the rest of the group, where $\a$ permutes $X^3$, $\beta$ permutes $X^5$ and $X^7$, and $\gamma$ permutes $X^1$. 
We choose to label each of the two 8 dimensional sets by $f_{\delta}\in\{1,2\}$, corresponding to either the case $X^5=X^7$ or $X^5\neq X^7$. In each of these sectors, we have an 8 dimensional representation $\mathcal{H}_\delta^{f_{\delta}}$ with basis states labelled by $X^1\in\{0,1/2\}$, and $X^3,X^5\in\{1/4,3/4\}$. The 8 states $\ket{i}_\delta^{f_{\delta}}=(X^1,X^3,X^5)$ are:
\begin{equation}
\begin{split}
    &\ket{1}_\delta^{f_{\delta}}=(0,1/4,1/4)\,,\quad\ket{2}_\delta^{f_{\delta}}=(0,1/4,3/4)\,,\quad\ket{3}_\delta^{f_{\delta}}=(0,3/4,1/4)\,,\\
    &\ket{4}_\delta^{f_{\delta}}=(0,3/4,3/4)\,,\quad\ket{5}_\delta^{f_{\delta}}=(1/2,1/4,1/4)\,,\quad\ket{6}_\delta^{f_{\delta}}=(1/2,1/4,3/4)\,,\\
    &\ket{7}_\delta^{f_{\delta}}=(1/2,3/4,1/4)\,,\quad\ket{8}_\delta^{f_{\delta}}=(1/2,3/4,3/4)\,\,.
\end{split}
\end{equation}
and the group action is:
\begin{gather}
    \a|_{\mathcal{H}_\delta^{f_{\delta}}}\,:\,\begin{pmatrix}\ket{1}_\delta^{f_{\delta}}\leftrightarrow\ket{5}_\delta^{f_{\delta}} \\ \ket{2}_\delta^{f_{\delta}}\leftrightarrow\ket{6}_\delta^{f_{\delta}} \\ \ket{3}_\delta^{f_{\delta}}\leftrightarrow\ket{7}_\delta^{f_{\delta}} \\ \ket{4}_\delta^{f_{\delta}}\leftrightarrow\ket{8}_\delta^{f_{\delta}}\end{pmatrix}\,,\quad \beta|_{\mathcal{H}_\delta^{f_{\delta}}}\,:\,\begin{pmatrix}\ket{1}_\delta^{f_{\delta}}\leftrightarrow\ket{2}_\delta^{f_{\delta}} \\ \ket{3}_\delta^{f_{\delta}}\leftrightarrow\ket{4}_\delta^{f_{\delta}} \\ \ket{5}_\delta^{f_{\delta}}\leftrightarrow\ket{6}_\delta^{f_{\delta}} \\ \ket{7}_\delta^{f_{\delta}}\leftrightarrow\ket{8}_\delta^{f_{\delta}}\end{pmatrix}\,,\quad\hbox{and}\quad
    \gamma|_{\mathcal{H}_\delta^{f_{\delta}}}\,:\,\begin{pmatrix}\ket{1}_\delta^{f_{\delta}}\leftrightarrow\ket{3}_\delta^{f_{\delta}} \\ \ket{2}_\delta^{f_{\delta}}\leftrightarrow\ket{4}_\delta^{f_{\delta}} \\ \ket{5}_\delta^{f_{\delta}}\leftrightarrow\ket{7}_\delta^{f_{\delta}} \\ \ket{6}_\delta^{f_{\delta}}\leftrightarrow\ket{8}_\delta^{f_{\delta}}\end{pmatrix}\,.
\end{gather}
The matrix representations are $8\times 8$ and can easily be constructed, and we find after imposing any commutation relations that any discrete torsion signs vanish.


\subsubsection*{$\gamma$ sector}

Here we get 16 fixed points with $X^1,X^5\in\{1/4,3/4\}$ and $X^2,X^6\in\{0,1/2\}$. In this sector, $\a$ and $\delta$ both permute $X^1$ and $\beta$ permutes $X^5$ - thus, the 16 states get grouped into 4 sets of 4 labelled by the choices of $X^2$ and $X^6$. Choosing a label $f_\gamma\in\{1,..,4\}$ to represent this choice, the 4 states within each sub sector correspond to the 2-tuples $(X^1,X^5)$:
\begin{equation}
\begin{split}
    &\ket{1}_\gamma^{f_{\gamma}}=(1/4,1/4)\,,\quad \ket{2}_\gamma^{f_{\gamma}}=(3/4,1/4)\,\,,\\
    &\ket{3}_\gamma^{f_{\gamma}}=(1/4,3/4)\,,\quad \ket{4}_\gamma^{f_{\gamma}}=(3/4,3/4)\,\,,
\end{split}
\end{equation}
with:
\begin{gather}
    \a|_{\mathcal{H}_\gamma^{f_\gamma}},\delta|_{\mathcal{H}_\gamma^{f_\gamma}}\,:\,\begin{pmatrix}\ket{1}_\gamma^{f_{\gamma}}\leftrightarrow\ket{2}_\gamma^{f_{\gamma}} \\ \ket{3}_\gamma^{f_{\gamma}}\leftrightarrow\ket{4}_\gamma^{f_{\gamma}}\end{pmatrix}\,,\quad\hbox{and}\quad \beta|_{\mathcal{H}_\gamma^{f_\gamma}}\,:\,\begin{pmatrix}\ket{1}_\gamma^{f_{\gamma}}\leftrightarrow\ket{3}_\gamma^{f_{\gamma}} \\ \ket{2}_\gamma^{f_{\gamma}}\leftrightarrow\ket{4}_\gamma^{f_{\gamma}}\end{pmatrix}\,\,.
\end{gather}
It is then clear that $\a\delta$ acts diagonally. We can then turn these into matrix representations, and after imposing any constraints we find:
\begin{equation}
    \a|_{\mathcal{H}_\gamma^{f_\gamma}}=\begin{pmatrix}H & 0 \\0 & H\end{pmatrix}\,,\quad\delta|_{\mathcal{H}_\gamma^{f_\gamma}}=\e_{f_\gamma}(\delta)\cdot \a|_{\mathcal{H}_\gamma^{f_\gamma}} \quad\hbox{and}\quad \beta|_{\mathcal{H}_\gamma^{f_\gamma}}=\begin{pmatrix}0 & \mathbb{I}_{2\times 2} \\ \mathbb{I}_{2\times 2} & 0\end{pmatrix}\,\,.
\end{equation}
So we get 4 sign degrees of freedom $\e_{f_\gamma}(\delta)$. Only the elements $\a\delta|_{\mathcal{H}_\gamma^{f_\gamma}}=\a\delta\gamma|_{\mathcal{H}_\gamma^{f_\gamma}}=\e_{f_\gamma}(\delta)\cdot\mathbb{I}_{4\times 4}$ act diagonally, and so we should look at these sectors next. 


\subsubsection*{$\alpha\delta$ and $\a\delta\gamma$}

The actions of $\a\delta$ and $\a\delta\gamma$ are:

\begin{center}
\begin{tabular}{ |c|c|c|c|c|c|c|c|c| } 
 \hline
  & $X^1$ & $X^2$ & $X^3$ & $X^4$ & $X^5$ & $X^6$ & $X^7$ & $X^8$\\ 
  \hline
  $\a\delta$ & + & - & $+\frac{1}{2}$ & - & $-\frac1{2}$ & + & $-\frac1{2}$ & +\\ 
  $\a\delta\gamma$ & $-\frac1{2}$ & + & $+\frac1{2}$ & - & + & - & $-\frac1{2}$ & + \\
 \hline
\end{tabular}
\end{center}

So, for $\a\delta$ we have fixed points in $X^2$, $X^4$, $X^5$ and $X^7$, and the action $X^3\rightarrow X^3+1/2$ means the winding number $n_3$ takes values in $\mathbb{Z}+1/2$. In the lowest energy state we must take $n_3=\pm 1/2$, and here $\a$ and $\delta$ both permute $n_3$, $\gamma$ acts diagonally, and $\beta$ permutes $X^5$ and $X^7$ (both $\in\{1/4,3/4\}$). This action groups the 32 states (16 fixed points/twist fields with two possible winding numbers each) into 8 sets of 4, with a label $f_{\a\delta}\in 1..8$ representing these sectors. Within each, we label states by $(X^5,n_3)$:
\begin{equation}
\begin{split}
    &\ket{1}_{\a\delta}^{f_{\a\delta}}=(1/4,+)\,,\quad \ket{3}_{\a\delta}^{f_{\a\delta}}=(3/4,+)\,\,,\\
    &\ket{2}_{\a\delta}^{f_{\a\delta}}=(1/4,-)\,,\quad \ket{4}_{\a\delta}^{f_{\a\delta}}=(3/4,-)\,\,.
\end{split}
\end{equation}
Here, $\pm$ is short for $n_3=\pm1/2$. So we find:
\begin{gather}
    \a|_{\mathcal{H}_{\a\delta}^{f_{\a\delta}}},\,\delta|_{\mathcal{H}_{\a\delta}^{f_{\a\delta}}}\,:\,\begin{pmatrix}\ket{1}_{\a\delta}^{f_{\a\delta}}\leftrightarrow\ket{2}_{\a\delta}^{f_{\a\delta}} \\ \ket{3}_{\a\delta}^{f_{\a\delta}}\leftrightarrow\ket{4}_{\a\delta}^{f_{\a\delta}}\end{pmatrix}\,,\quad\hbox{and}\quad \beta|_{\mathcal{H}_{\a\delta}^{f_{\a\delta}}}\,:\,\begin{pmatrix}\ket{1}_{\a\delta}^{f_{\a\delta}}\leftrightarrow\ket{3}_{\a\delta}^{f_{\a\delta}} \\ \ket{2}_{\a\delta}^{f_{\a\delta}}\leftrightarrow\ket{4}_{\a\delta}^{f_{\a\delta}}\end{pmatrix}\,\,,
\end{gather}
with $\gamma$ diagonal. After imposing representation constraints, only the generator $\gamma$ retains a phase and has a matrix representation of $\gamma=\e_{f_{\a\delta}}(\gamma)\cdot\mathbb{I}_{4\times 4}$. As required, the discrete torsion sign is independent of the winding number, and the 16 sign degrees of freedom get organized across the twist fields into 8 sets of 2.

For $\a\gamma\delta$, the twist fields have labels $(X^1,X^4,X^6,X^7,n_3)$ with $X^4,X^6\in\{0,1/2\}$, $X^1,X^7\in\{1/4,3/4\}$ and $n_3\in \{1/2,-1/2\}$. $\a$ and $\delta$ permute $X^1$, $X^3$ and $n_3$, while $\beta$ permutes $X^7$ and $\gamma$ acts diagonally. This orders the 32 states into 4 sets of 8, labelled by a number $f_{\a\delta\gamma}\in\{1,..,4\}$ and by $(X^1,X^7,n_3)$ within each set:
\begin{equation}
\begin{split}
    &\ket{1}_{\a\delta\gamma}^{f_{\a\delta\gamma}}=(1/4,1/4,+)\,,\quad\ket{2}_{\a\delta\gamma}^{f_{\a\delta\gamma}}=(1/4,3/4,+)\,,\quad\ket{3}_{\a\delta\gamma}^{f_{\a\delta\gamma}}=(3/4,1/4,+)\,,\\
    &\ket{4}_{\a\delta\gamma}^{f_{\a\delta\gamma}}=(3/4,3/4,+)\,,\quad\ket{5}_{\a\delta\gamma}^{f_{\a\delta\gamma}}=(1/4,1/4,-)\,,\quad\ket{6}_{\a\delta\gamma}^{f_{\a\delta\gamma}}=(1/4,3/4,-)\,,\\
    &\ket{7}_{\a\delta\gamma}^{f_{\a\delta\gamma}}=(3/4,1/4,-)\,,\quad\ket{8}_{\a\delta\gamma}^{f_{\a\delta\gamma}}=(3/4,3/4,-)\,\,,
\end{split}
\end{equation}
The group action is:
\begin{gather}
    \a|_{\mathcal{H}_{\a\delta\gamma}^{f_{\a\delta\gamma}}},\,\delta|_{\mathcal{H}_{\a\delta\gamma}^{f_{\a\delta\gamma}}}\,:\,\begin{pmatrix}\ket{1}_{\a\delta\gamma}^{f_{\a\delta\gamma}}\leftrightarrow\ket{7}_{\a\delta\gamma}^{f_{\a\delta\gamma}} \\ \ket{2}_{\a\delta\gamma}^{f_{\a\delta\gamma}}\leftrightarrow\ket{8}_{\a\delta\gamma}^{f_{\a\delta\gamma}} \\ \ket{3}_{\a\delta\gamma}^{f_{\a\delta\gamma}}\leftrightarrow\ket{5}_{\a\delta\gamma}^{f_{\a\delta\gamma}} \\ \ket{4}_{\a\delta\gamma}^{f_{\a\delta\gamma}}\leftrightarrow\ket{6}_{\a\delta\gamma}^{f_{\a\delta\gamma}}\end{pmatrix}\,,\quad\hbox{and}\quad\beta|_{\mathcal{H}_{\a\delta\gamma}^{f_{\a\delta\gamma}}}\,:\,\begin{pmatrix}\ket{1}_{\a\delta\gamma}^{f_{\a\delta\gamma}}\leftrightarrow\ket{2}_{\a\delta\gamma}^{f_{\a\delta\gamma}} \\ \ket{3}_{\a\delta\gamma}^{f_{\a\delta\gamma}}\leftrightarrow\ket{4}_{\a\delta\gamma}^{f_{\a\delta\gamma}} \\ \ket{5}_{\a\delta\gamma}^{f_{\a\delta\gamma}}\leftrightarrow\ket{6}_{\a\delta\gamma}^{f_{\a\delta\gamma}} \\ \ket{7}_{\a\delta\gamma}^{f_{\a\delta\gamma}}\leftrightarrow\ket{8}_{\a\delta\gamma}^{f_{\a\delta\gamma}}\end{pmatrix}\,\,.
\end{gather}
After organizing the phases, we find that $\delta=\e_{f_{\a\gamma\delta}}(\delta,\gamma)\cdot\a$, $\gamma=\e_{f_{\a\gamma\delta}}(\delta,\gamma)\cdot \mathbb{I}_{8\times 8}$ and $\beta$ has no phases (again, $\e_{f_{\a\gamma\delta}}(\delta,\gamma)$ emphasizes that both $\gamma$ and $\delta$ get signs in this sector). This way, $\a\delta\gamma|_{\mathcal{H}_{\a\delta\gamma}^{f_{\a\delta\gamma}}}=\mathbb{I}_{8\times 8}$. Once again the discrete torsion phases are winding independent, and the 16 signs get organized into 8 sets of 2 across the twist fields labelled by $f_{\a\gamma\delta}$.


\subsubsection*{$\alpha\gamma$, $\beta\gamma$, $\delta\gamma$, $\beta\delta$, $\a\beta\gamma$, $\a\beta\delta$ and $\a\beta\delta\gamma$}

For the rest of the composite group elements, the analysis follows the above structure and we find that imposing commutation constraints or equalities such as $\a\gamma|_{\mathcal{H}_{\a\gamma}^{f_{\a\gamma}}}{\stackrel{!}{=}}\mathbb{I}$ allow us to absorb all discrete torsion phases. So, no new constraints will arise here.


\subsection*{Constraints}

In summary, we have found that their are discrete torsion phases in the $\a$, $\beta$, $\gamma$, $\a\beta$, $\a\delta$ and $\a\delta\gamma$ sectors. We would now like to understand how modular invariance relates these phases.

First, let us consider the $\gamma$, $\a\delta$ and $\a\delta\gamma$ twisted sectors. After taking into account the identification of discrete torsion phases, we find:
\begin{equation}
\begin{split}
    &Z_{\gamma;\a\delta}(-1/\tau)=Z_{\a\delta;\gamma}(\tau)\Rightarrow4\sum_{f_{\gamma}=1}^4\e_{f_{\gamma}}(\delta)=2\sum_{f_{\a\delta}=1}^8\e_{f_{\a\delta}}(\gamma)\\
    &Z_{\gamma;\a\delta\gamma}(-1/\tau)=Z_{\a\delta\gamma;\gamma}(\tau)\Rightarrow 4\sum_{f_{\gamma}=1}^4\e_{f_\gamma}(\delta)=4\sum_{f_{\a\delta\gamma}=1}^4\e_{f_{\a\delta\gamma}}(\delta,\gamma)
\end{split}\,\,.
\end{equation}
As a result, after a possible reordering of the labels $f_g$ we may set $f_\gamma=f_{\a\delta\gamma}=f_{\a\delta}$ and get:
\begin{equation}
    \e_{f_\gamma}\equiv \e_{f_\gamma}(\delta)= \e_{f_{\gamma}}(\delta,\gamma)=\e_{f_{\gamma}+4}(\gamma)\,\,.
\end{equation}

Now consider the case of $\a$, $\beta$ and $\a\beta$. For $\a$ and $\beta$ we find:
\begin{equation}
    4\sum_{f_\a=1}^4\e_{f_\a}(\beta)=4\sum_{f_\beta=1}^4\e_{f_\beta}(\alpha)\,\,.
\end{equation}
So, after setting $f_{\a}=f_{\beta}$:
\begin{equation}
    \e_{f_\alpha}\equiv\e_{f_\a}(\beta)=\e_{f_\a}(\a)\,\,.
    \label{eq:fafb}
\end{equation}

However, consider the case of $\a\beta$. Taking into account the identification of phases across the twist fields, we find:
\begin{equation*}
    4\sum_{f_{\a\beta}=1}^{64}\e_{f_{\a\beta}}(\a,\beta)=16\left(4\sum_{f_\a=1}^{4}\e_{f_\a}\right) \Rightarrow \sum_{f_{\a\beta}=1}^{64}\e_{f_{\a\beta}}(\a,\beta)=16\sum_{f_\a=1}^{4}\e_{f_\a}\,\,,
\end{equation*}
where we also used equation \eqref{eq:fafb}. Thus, we may set:
\begin{equation}
    \e_{f_\a}=\e_{f_{\a\beta}+4k}(\a,\beta)\,\,,
\end{equation}
for $k=0,\,...,\,15$ and $f_{\a}=1,\,...,\,4$.

\providecommand{\href}[2]{#2}\begingroup\raggedright\endgroup


\begin{thebibliography}{10}

\bibitem{Shatashvili:1994zw}
S.~L. Shatashvili and C.~Vafa, ``{Superstrings and manifold of exceptional
  holonomy},'' \href{http://dx.doi.org/10.1007/BF01671569}{{\em Selecta Math.}
  {\bfseries 1} (1995) 347},
\href{http://arxiv.org/abs/hep-th/9407025}{{\ttfamily arXiv:hep-th/9407025
  [hep-th]}}.

\bibitem{Figueroa-OFarrill:1996tnk}
J.~M. Figueroa-O'Farrill, ``{A Note on the extended superconformal algebras
  associated with manifolds of exceptional holonomy},''
  \href{http://dx.doi.org/10.1016/S0370-2693(96)01506-7}{{\em Phys. Lett.}
  {\bfseries B392} (1997) 77--84},
\href{http://arxiv.org/abs/hep-th/9609113}{{\ttfamily arXiv:hep-th/9609113
  [hep-th]}}.

\bibitem{joyce1996I}
D.~D. Joyce, ``Compact riemannian 7-manifolds with holonomy $g_2$. i,'' {\em J.
  Differential Geom.} {\bfseries 43} no.~2, (1996) 291--328.

\bibitem{joyce1996II}
D.~D. Joyce, ``Compact riemannian 7-manifolds with holonomy $g\sb 2$. ii,''
  {\em J. Differential Geom.} {\bfseries 43} no.~2, (1996) 329--375.

\bibitem{joyce1996spin7}
D.~D. Joyce, ``Compact 8-manifolds with holonomy spin(7),'' {\em Inventiones
  mathematica} {\bfseries 123} (1996) 507--552.

\bibitem{Acharya:1997rh}
B.~S. Acharya, ``{On mirror symmetry for manifolds of exceptional holonomy},''
  \href{http://dx.doi.org/10.1016/S0550-3213(98)00140-0}{{\em Nucl. Phys.}
  {\bfseries B524} (1998) 269--282},
\href{http://arxiv.org/abs/hep-th/9707186}{{\ttfamily arXiv:hep-th/9707186
  [hep-th]}}.

\bibitem{Acharya:1996fx}
B.~S. Acharya, ``{Dirichlet Joyce manifolds, discrete torsion and duality},''
  \href{http://dx.doi.org/10.1016/S0550-3213(97)00163-6}{{\em Nucl. Phys.}
  {\bfseries B492} (1997) 591--606},
\href{http://arxiv.org/abs/hep-th/9611036}{{\ttfamily arXiv:hep-th/9611036
  [hep-th]}}.

\bibitem{Strominger:1996it}
A.~Strominger, S.-T. Yau, and E.~Zaslow, ``{Mirror symmetry is T duality},''
  \href{http://dx.doi.org/10.1016/0550-3213(96)00434-8}{{\em Nucl. Phys.}
  {\bfseries B479} (1996) 243--259},
\href{http://arxiv.org/abs/hep-th/9606040}{{\ttfamily arXiv:hep-th/9606040
  [hep-th]}}.

\bibitem{Gaberdiel:2004vx}
M.~R. Gaberdiel and P.~Kaste, ``{Generalized discrete torsion and mirror
  symmetry for g(2) manifolds},''
  \href{http://dx.doi.org/10.1088/1126-6708/2004/08/001}{{\em JHEP} {\bfseries
  08} (2004) 001},
\href{http://arxiv.org/abs/hep-th/0401125}{{\ttfamily arXiv:hep-th/0401125
  [hep-th]}}.

\bibitem{Chuang:2004th}
W.-y. Chuang, ``{A Note on mirror symmetry for manifolds with spin(7)
  holonomy},'' \href{http://dx.doi.org/10.1088/1751-8113/43/23/235403}{{\em J.
  Phys.} {\bfseries A43} (2010) 235403},
\href{http://arxiv.org/abs/hep-th/0406151}{{\ttfamily arXiv:hep-th/0406151
  [hep-th]}}.

\bibitem{Papadopoulos:1995da}
G.~Papadopoulos and P.~K. Townsend, ``{Compactification of D = 11 supergravity
  on spaces of exceptional holonomy},''
  \href{http://dx.doi.org/10.1016/0370-2693(95)00929-F}{{\em Phys. Lett.}
  {\bfseries B357} (1995) 300--306},
\href{http://arxiv.org/abs/hep-th/9506150}{{\ttfamily arXiv:hep-th/9506150
  [hep-th]}}.

\bibitem{MR2024648}
A.~Kovalev, ``Twisted connected sums and special {R}iemannian holonomy,''
  \href{http://dx.doi.org/10.1515/crll.2003.097}{{\em J. Reine Angew. Math.}
  {\bfseries 565} (2003) 125--160}.
  \url{http://dx.doi.org/10.1515/crll.2003.097}.

\bibitem{MR3109862}
A.~Corti, M.~Haskins, J.~Nordstr{\"o}m, and T.~Pacini, ``Asymptotically
  cylindrical {C}alabi-{Y}au 3-folds from weak {F}ano 3-folds,''
  \href{http://dx.doi.org/10.2140/gt.2013.17.1955}{{\em Geom. Topol.}
  {\bfseries 17} no.~4, (2013) 1955--2059}.
  \url{http://dx.doi.org/10.2140/gt.2013.17.1955}.

\bibitem{Corti:2012kd}
A.~Corti, M.~Haskins, J.~Nordstr{\"o}m, and T.~Pacini,
  ``{$\mathrm{G}_{2}$-manifolds and associative submanifolds via semi-Fano
  $3$-folds},'' \href{http://dx.doi.org/10.1215/00127094-3120743}{{\em Duke
  Math. J.} {\bfseries 164} no.~10, (2015) 1971--2092},
\href{http://arxiv.org/abs/1207.4470}{{\ttfamily arXiv:1207.4470 [math.DG]}}.

\bibitem{Braun:2017ryx}
A.~P. Braun and M.~Del~Zotto, ``{Mirror Symmetry for $G_2$-Manifolds: Twisted
  Connected Sums and Dual Tops},''
  \href{http://dx.doi.org/10.1007/JHEP05(2017)080}{{\em JHEP} {\bfseries 05}
  (2017) 080},
\href{http://arxiv.org/abs/1701.05202}{{\ttfamily arXiv:1701.05202 [hep-th]}}.

\bibitem{Braun:2017csz}
A.~P. Braun and M.~Del~Zotto, ``{Towards Generalized Mirror Symmetry for
  Twisted Connected Sum $G_2$ Manifolds},''
  \href{http://dx.doi.org/10.1007/JHEP03(2018)082}{{\em JHEP} {\bfseries 03}
  (2018) 082},
\href{http://arxiv.org/abs/1712.06571}{{\ttfamily arXiv:1712.06571 [hep-th]}}.

\bibitem{joyce2000compact}
D.~Joyce, {\em Compact Manifolds with Special Holonomy}.
\newblock Oxford mathematical monographs. Oxford University Press, 2000.

\bibitem{Batyrev:1994hm}
V.~V. Batyrev, ``{Dual polyhedra and mirror symmetry for Calabi-Yau
  hypersurfaces in toric varieties},'' {\em J. Alg. Geom.} {\bfseries 3} (1994)
  493--545,
\href{http://arxiv.org/abs/alg-geom/9310003}{{\ttfamily arXiv:alg-geom/9310003
  [alg-geom]}}.

\bibitem{Braun:2016igl}
A.~P. Braun, ``{Tops as building blocks for G$_{2}$ manifolds},''
  \href{http://dx.doi.org/10.1007/JHEP10(2017)083}{{\em JHEP} {\bfseries 10}
  (2017) 083},
\href{http://arxiv.org/abs/1602.03521}{{\ttfamily arXiv:1602.03521 [hep-th]}}.

\bibitem{Aspinwall:1994rg}
P.~S. Aspinwall and D.~R. Morrison, ``{String theory on K3 surfaces},''
\href{http://arxiv.org/abs/hep-th/9404151}{{\ttfamily arXiv:hep-th/9404151
  [hep-th]}}.

\bibitem{Braun:2017uku}
A.~P. Braun and S.~Schafer-Nameki, ``{Compact, Singular G2-Holonomy Manifolds
  and M/Heterotic/F-Theory Duality},''
\href{http://arxiv.org/abs/1708.07215}{{\ttfamily arXiv:1708.07215 [hep-th]}}.

\bibitem{Halverson:2014tya}
J.~Halverson and D.~R. Morrison, ``{The landscape of M-theory compactifications
  on seven-manifolds with G$_{2}$ holonomy},''
  \href{http://dx.doi.org/10.1007/JHEP04(2015)047}{{\em JHEP} {\bfseries 04}
  (2015) 047},
\href{http://arxiv.org/abs/1412.4123}{{\ttfamily arXiv:1412.4123 [hep-th]}}.

\bibitem{Halverson:2015vta}
J.~Halverson and D.~R. Morrison, ``{On gauge enhancement and singular limits in
  G$_{2}$ compactifications of M-theory},''
  \href{http://dx.doi.org/10.1007/JHEP04(2016)100}{{\em JHEP} {\bfseries 04}
  (2016) 100},
\href{http://arxiv.org/abs/1507.05965}{{\ttfamily arXiv:1507.05965 [hep-th]}}.

\bibitem{Guio:2017zfn}
T.~C. d.~C. Guio, H.~Jockers, A.~Klemm, and H.-Y. Yeh, ``{Effective action from
  M-theory on twisted connected sum $G_2$-manifolds},''
\href{http://arxiv.org/abs/1702.05435}{{\ttfamily arXiv:1702.05435 [hep-th]}}.

\bibitem{Braun:2018fdp}
A.~P. Braun, M.~Del~Zotto, J.~Halverson, M.~Larfors, D.~R. Morrison, and
  S.~Schafer-Nameki, ``{Infinitely Many M2-instanton Corrections to M-theory on
  $G_2$-manifolds},''
\href{http://arxiv.org/abs/1803.02343}{{\ttfamily arXiv:1803.02343 [hep-th]}}.

\bibitem{Acharya:2018nbo}
B.~S. Acharya, A.~P. Braun, E.~E. Svanes, and R.~Valandro, ``{Counting
  Associatives in Compact $G_2$ Orbifolds},''
\href{http://arxiv.org/abs/1812.04008}{{\ttfamily arXiv:1812.04008 [hep-th]}}.

\bibitem{Braun:2018joh}
A.~P. Braun and S.~Schäfer-Nameki, ``{Spin(7)-manifolds as generalized
  connected sums and 3d $\mathcal{N}=1$ theories},''
  \href{http://dx.doi.org/10.1007/JHEP06(2018)103}{{\em JHEP} {\bfseries 06}
  (2018) 103},
\href{http://arxiv.org/abs/1803.10755}{{\ttfamily arXiv:1803.10755 [hep-th]}}.

\bibitem{Fiset:2018huv}
M.-A. Fiset, ``{Superconformal algebras for twisted connected sums and G$_{2}$
  mirror symmetry},'' \href{http://dx.doi.org/10.1007/JHEP12(2018)011}{{\em
  JHEP} {\bfseries 12} (2018) 011},
\href{http://arxiv.org/abs/1809.06376}{{\ttfamily arXiv:1809.06376 [hep-th]}}.

\bibitem{vafa1986modular}
C.~Vafa, ``Modular invariance and discrete torsion on orbifolds,'' {\em Nuclear
  Physics B} {\bfseries 273} no.~3-4, (1986) 592--606.

\bibitem{Witten:1982im}
E.~Witten, ``{Supersymmetry and Morse theory},''
{\em J. Diff. Geom.} {\bfseries 17} no.~4, (1982) 661--692.

\bibitem{Aspinwall:1995rb}
P.~S. Aspinwall, D.~R. Morrison, and M.~Gross, ``{Stable singularities in
  string theory},'' \href{http://dx.doi.org/10.1007/BF02104911}{{\em Commun.
  Math. Phys.} {\bfseries 178} (1996) 115--134},
\href{http://arxiv.org/abs/hep-th/9503208}{{\ttfamily arXiv:hep-th/9503208
  [hep-th]}}.

\bibitem{1998math......9072G}
M.~{Gross}, ``{Special Lagrangian Fibrations II: Geometry},'' {\em ArXiv
  Mathematics e-prints} (1998) ,
  \href{http://arxiv.org/abs/math/9809072}{{\ttfamily math/9809072}}.

\bibitem{2015arXiv150502734C}
D.~{Crowley}, S.~{Goette}, and J.~{Nordstr{\"o}m}, ``{An analytic invariant of
  G\_2 manifolds},'' {\em ArXiv e-prints} (May, 2015) ,
  \href{http://arxiv.org/abs/1505.02734}{{\ttfamily arXiv:1505.02734
  [math.GT]}}.

\bibitem{Vafa:1994rv}
C.~Vafa and E.~Witten, ``{On orbifolds with discrete torsion},''
  \href{http://dx.doi.org/10.1016/0393-0440(94)00048-9}{{\em J. Geom. Phys.}
  {\bfseries 15} (1995) 189--214},
\href{http://arxiv.org/abs/hep-th/9409188}{{\ttfamily arXiv:hep-th/9409188
  [hep-th]}}.

\bibitem{joyce1996spin7_new}
D.~D. Joyce, ``A new construction of compact 8-manifolds with holonomy
  spin(7),'' {\em J. Differential Geom.} {\bfseries 53} (1999) 89--130,
  \href{http://arxiv.org/abs/9910002}{{\ttfamily arXiv:9910002 [math.DG]}}.

\bibitem{2017arXiv170709325J}
D.~{Joyce} and S.~{Karigiannis}, ``{A new construction of compact torsion-free
  \$G\_2\$-manifolds by gluing families of Eguchi-Hanson spaces},'' {\em arXiv
  e-prints} (Jul, 2017) arXiv:1707.09325,
  \href{http://arxiv.org/abs/1707.09325}{{\ttfamily arXiv:1707.09325
  [math.DG]}}.

\bibitem{Gopakumar:1996mu}
R.~Gopakumar and S.~Mukhi, ``{Orbifold and orientifold compactifications of F -
  theory and M - theory to six-dimensions and four-dimensions},''
  \href{http://dx.doi.org/10.1016/0550-3213(96)00460-9}{{\em Nucl. Phys.}
  {\bfseries B479} (1996) 260--284},
\href{http://arxiv.org/abs/hep-th/9607057}{{\ttfamily arXiv:hep-th/9607057
  [hep-th]}}.

\end{thebibliography}
\end{document}